\documentclass{jfm}
\usepackage{graphicx}
\usepackage{bm}
\usepackage{dcolumn}
\usepackage{amsmath,amssymb}
 \usepackage{color}
\usepackage[binary-units=true]{siunitx}
\usepackage{systeme,mathtools}
\usepackage{hyperref}
\usepackage{booktabs}

\DeclareSIUnit\atm{atm}

\newcommand{\ii}{\textrm{i}}
\newcommand{\ee}{\textrm{e}}
\newcommand{\dd}{\textrm{d}}
\newcommand{\bigO}{\textrm{O}}
\newcommand{\grad}{\boldsymbol{\nabla}}

\newcommand{\Mach}{\mathit{M\kern-2pt a}\kern 1pt} 
\newcommand{\Ca}{\mathit{C\kern-2pt a}\kern 1pt} 
\DeclareMathOperator{\re}{Re}
\DeclareMathOperator{\im}{Im}

\renewcommand{\vec}[1]{\bm{#1}}
\newcommand{\diverg}[1]{\grad\cdot{#1}}

\hypersetup{
    colorlinks=true,
    linkcolor=blue,
    filecolor=cyan,      
    urlcolor=cyan,
    citecolor = blue
    }

\shorttitle{Cylindrical vortex normal modes}
\shortauthor{Marques \& Silva}

\title{Vortex Modes in Acoustofluidic Cylindrical Resonators}

\author{Alisson S. Marques and Glauber T. Silva
\corresp{\email{gtomaz@fis.ufal.br} } 
}
 
\affiliation{
Group of
Physical Acoustics \& Microfluidics, Federal University of Alagoas, Macei\'o, AL 57072-970, Brazil}

\begin{document}

\maketitle

\begin{abstract}
This paper presents a theoretical investigation of vortex modes in acoustofluidic cylindrical resonators with rigid boundaries and viscous fluids. By solving the Helmholtz equation for linear pressure, incorporating boundary conditions that account for no-slip surfaces and vortex and nonvortex excitation at the base, we analyze both single- and dual-eigenfunction modes near system resonance. 
The results demonstrate that single vortex modes generate spin angular momentum exclusively along the axial direction, while dual modes introduce a transverse spin component due to the nonlinear interaction between axial and transverse ultrasonic waves, even in the absence of vortex excitation. 
We find that nonlinear acoustic fields, including energy density, radiation force potential, and spin, scale with the square of the shear wave number, defined as the ratio of the cavity radius to the boundary layer depth. Theoretical predictions align closely with finite element simulations based on a thermoviscous model for an acoustofluidic cavity with adiabatic and rigid walls. 
These findings hold particular significance for acoustofluidic systems, offering potential applications in the precise control of cells and microparticles.
\end{abstract}


\section{Introduction}

The control of acoustic waves within confined geometries has led to remarkable advancements across various scientific and technological domains. 
Cylindrical resonator cavities, characterized by their closed, rigid walls and fluid-filled interiors, serve as fundamental platforms for exploring intricate acoustic phenomena. 
\cite{Kundt1874} were the first to describe the use of acoustic standing waves to gather initially uniformly distributed particles at specific planes of acoustic pressure, either nodal or antinodal. 
More recently, a renewed interest in studying wave dynamics within  cylindrical resonators has grown significantly due to their potential to manipulate microparticles and cells within acoustofluidic microcavities \citep{Coakley1997,Spengler2000,Goddard2005,Lenshof2012,Wang2012,Gralinski2014,Gutierrez-Ramos2018,Santos2021,Rocha2023,Fuchsluger2024}.
In turn, acoustofluidics uses the forces of ultrasonic waves to manipulate and control cell or micro/nanoparticles within microfluidic structures.

One of the earliest theoretical explorations into cylindrical resonators can be credited to \citet[p. 49-68]{Rayleigh1945}. This seminal work later became a fundamental part of textbooks, as seen in~\citet[p. 381-431]{Morse1948}. 
These studies extensively employed the classical wave equation, formulated in terms of acoustic pressure only, and its time-harmonic equivalent, the Helmholtz equation. However, these existing models have overlooked the impact of losses within the boundary layer adjacent to the cylindrical cavity walls. 
The wave dynamics, incorporating boundary layer effects under the isentropic hypothesis (where entropy remains constant with heat transfer being negligible), encompass two distinct spatial scales: the acoustic wavelength ($\lambda$) and the depth of the boundary layer ($\delta$), where $\lambda\gg \delta$. Some recent efforts have been dedicated to deriving boundary conditions for acoustic pressure in cavities with curved boundaries, where the radius of curvature is much larger than the depth of the viscous boundary layer, primarily motivated by numerical simulation purposes~\citep{Berggren2023,Bach2018}.

We focus our investigation on cylindrical cavity resonators similar to those used in acoustofluidic applications for manipulating cells and microparticles. 
This manipulation is driven by ultrasonic forces and torques, namely,
the acoustic radiation force~\citep{Doinikov2003,Baudoin2019b} and torque~\citep{Zhang2011c,Silva2012,Gong2019b}, and Stokes drag by acoustic streaming~\citep{Rednikov2011,Pavlic2021}. The radiation force concentrates suspended particles within pressure nodes or antinodes of a standing wave, depending on their density and compressibility compared with the surrounding fluid~\citep{Gorkov1962}. 
The radiation torque may induce  rotation around the particle principal axis due to spin angular momentum transfer from the wave to the particle, which is referred to as \textit{the spin-torque effect}~\citep{Silva2014, Zhang2014, Mitri2016,Bliokh2019,Lopes2020}. 
Drag forces can be caused by acoustic streaming, which represented by sustained flow rolls. 
However, polystyrene particles with a diameter larger than $\SI{1.4}{\micro\meter}$ in water, under the action of a $\SI{2}{\mega\hertz}$ ultrasonic wave,  are more responsive to the acoustic radiation force~\citep{Barnkob2012}. 
Given the focus on micro-sized particles and cells above this critical diameter, our analysis is solely devoted to scenarios involving acoustic radiation force and spin angular momentum.

To investigate the mentioned acoustofluidic phenomena in rigid cylindrical cavities, we employ a perturbation method in the fluid conservation equations with the adiabatic assumption to obtain the Helmholtz equation for the linear acoustic pressure. 
The boundary conditions at the walls deviate from the conventional Neumann condition, typically used to describe lossless wave phenomena, by a perturbation of the order of the viscous boundary layer depth divided by the wavelength, $\delta/\lambda$. 
The resultant boundary-value problem for the acoustic pressure is solved for a vortex excitation applied at the cavity's base, giving rise to either a single- or dual-frequency mode (the occurrence of two single modes simultaneously) configuration in the resonator. 
While the single modes depend only on frequency selection, the dual modes are also related to the cavity's aspect ratio.

The developed theory is showcased for an acoustofluidic microcavity resonator excited by sinusoidal vibration at the bottom.
We found that near resonance, the amplitude of the nonlinear effects for decoupled axial and vortex modes scales with the square of the shear wave number~\citep{Tijdeman1975}, $\textit{Sh}^2=(R/\delta)^2$, where $R$ is the cavity radius. 
Moreover, the findings show that single vortex modes produce spin angular momentum only in the axial direction, whereas dual modes create an additional transverse spin component as a result of the nonlinear interaction between axial and transverse ultrasonic waves, even without vortex excitation.
Finally, the obtained results are compared with those computed with finite element method using the linear thermoviscous model with rigid and adiabatic walls. 
Excellent agreement is found between the computational simulations and proposed theory.
To conclude, by incorporating effects of the viscous boundary layer and offering a detailed examination of the nonlinear acoustic phenomena within cylindrical resonators, our work extends the existing lossless theory~~\citep{Barmatz1985,Xu2019,Leao-Neto2021} and provides a robust basis for future numerical and experimental  studies.

\section{Physical model}

\subsection{Fluid dynamics equations}
        Consider a cylindrical microcavity resonator of radius $R$ and height $H$
        filled with a quiescent viscous liquid.
        The rigid boundary walls that enclose the cavity are represented by $S=S_\text{T}\cup S_\text{B}\cup S_\text{L}$, with $S_\text{T}$, $S_\text{B}$, and $S_\text{L}$ denoting the top, bottom, and lateral boundary, respectively.
        The Eulerian reference frame is placed in the center of $S_\text{B}$ at rest.
        The fluid flow is described by a mass density $\rho$, pressure $p$, and fluid velocity $\tilde{\vec{v}}$ as a function of position $\vec{r}$ and time $t$.
        To take advantage of the cavity's symmetry,  cylindrical coordinates is adopted $(\varrho, \varphi,z)$, with position vector $\vec{r} = \varrho\, \vec{e}_\varrho + \varphi\, \vec{e}_\varphi + z\, \vec{e}_z$, in which $\vec{e}_i$ $(i=\varrho,\varphi,z)$ represents the transverse radial, azimuthal, and axial unit vectors.
        % \begin{figure}
        % \begin{center}
        %     \includegraphics[scale=.5]{cavity.pdf} 
        % \end{center}
        % \caption{Cylindrical cavity resonator of height $H$ and radius $R$.
        % A Cartesian coordinate system is set in the center of the cavity's bottom.
        % \label{fig:resonator}}
        % \end{figure}
        
        The unperturbed state of the liquid is described by an ambient density $\rho_0$, pressure $p_0$, adiabatic speed of sound $c_0$, isentropic compressibility $\kappa_0=1/\rho_0 c_0^2$, and dynamic $\eta_0$ and bulk $\zeta_0$ viscosity.
        These parameters remain constant for the analysis under consideration.
     The physical parameters of water are listed in table~\ref{tab:PhysicalProperties}.

    In the investigation of acoustic wave propagation within the enclosed cavity, the process is assumed to be isentropic, with both internal heat generation and conduction effects being considered negligible. 
    This assumptions allows for  a simplification in complex interactions between thermodynamic and acoustic phenomena. A detailed analysis of adopting the isentropic hypothesis is presented in appendix~\ref{app:isentropic}. 

    The fluid dynamics equations are described by the mass and linear momentum conservation equations along with the density-pressure constitutive relation, 
        \begin{subequations}
        \label{acoustofluidic_equations}
            \begin{align}
            \label{EqDensity}
                \partial_t \rho
                + \grad \cdot (\rho {\vec{v}}) &= 0,\\
                \label{EqVelocity}
                    \partial_t (\rho {\vec{v}})
                    -\grad \cdot \left(\vec{\sigma} - \rho {\vec{v}}{\vec{v}} \right)
                    &= \vec{0},\\
                p &= c_0^2 \rho.
                \label{state_equation}
            \end{align}
        \end{subequations}
        The stress tensor is given by
    \begin{equation}
        \vec{\sigma} = 
        \left[- p + \left(
        \zeta_0 -\tfrac{2}{3}\eta_0
        \right) (\grad \cdot {\vec{v}})\right]
        \mathsfbi{I} 
        + \eta_0\! \left[\grad {\vec{v}} + \left(\grad {\vec{v}}\right)^\text{T}
        \right],
    \end{equation}
    where $\mathsfbi{I}$ is the unit tensor and the superscript $^\text{T}$ denotes the transpose of a tensor.
    The energy conservation equation is not needed to describe acoustic phenomenon within the isentropic hypothesis. 
\begin{table}
\centering
\begin{tabular}{l S[table-format=4.3] r }
Physical property & {Water} & {Unit} \\
\vspace{-.2cm}
\\
Density ($\rho_0$) & 997.00  & \si{\kilogram\per\cubic\meter} \\
Shear viscosity ($\eta_0$)& 0.88 & \si{\milli\pascal\second} \\
Bulk viscosity ($\zeta_0$)& 2.47 & \si{\milli\pascal\second} \\
Speed of sound ($c_0$)& 1496.70 &\si{\meter\per\second} \\
Isentropic compressibility ($\kappa_0$)& 0.45 &  \si{\per\giga\pascal} \\
Thermal diffusivity ($D_{\text{th},0}$) &  0.15 
& \si{\milli\meter\squared\per\second}
\end{tabular}
\caption{Physical properties of water at \SI{25}{\degreeCelsius}~\citep{Holmes2011}.
\label{tab:PhysicalProperties} }
\end{table}

\subsection{Perturbation expansion}
An acoustic wave can be injected into
the microcavity through a time-harmonic oscillation with  angular frequency $\omega$, at the bottom surface $S_\text{B}$ at $z=0$.
By considering that $S_\text{B}$ moves back and forth along the $z$ axis, we express the surface displacement as
\begin{equation}
        s(\varrho, \varphi; t) =
        d(\varrho, \varphi)\,
        \ee^{-\ii \omega t + \ii \tfrac{\upi}{2}},
        \quad \varrho < R,  
\end{equation}
where $d$ is the displacement function.
The phase $\upi/2$ is added here for convenience.
Note that no lateral motion of $S_\text{B}$ will be taken into account in our analysis.
The corresponding excitation velocity is 
\begin{equation}
    \label{v1ex}
    v_\text{ex}(\varrho,\varphi; t) =\partial_t{s} =  \omega\, d(\varrho,\varphi)\, 
    \ee^{-\ii \omega t}. 
\end{equation}

In the present analysis, the peak velocity of $S_\text{B}$ is assumed to be much smaller than the speed of sound in the medium.
In terms of the Mach number, this implies $\Mach=  \omega d_0/c_0\ll 1$ in which
$d_0$ is the given peak displacement at $S_\text{B}$.
To solve the fluid dynamics equations given in~\eqref{acoustofluidic_equations} in this approximation, the fields are conveniently expanded as
\begin{subequations}
    \label{expansions}
    \begin{align}
        \label{densityMach}
        \rho(\vec{r},t) &= \rho_0\left[1 + \Mach  \,\tilde{\rho}_1(\vec{r}) \ee^{-\ii \omega t}\right],\\
        p(\vec{r},t) &= p_0 + \Mach  \kappa_0^{-1}\, \tilde{p}_1(\vec{r}) \ee^{-\ii \omega t},\\
        \tilde{\vec{v}}(\vec{r},t) &= \vec{0} + \Mach c_0 \,\tilde{\vec{v}}_1(\vec{r}) \ee^{-\ii \omega t}.
        \label{v1Mach}
    \end{align}
\end{subequations}
The dimensionless amplitude functions $\tilde{p}_1$, $\tilde{\rho}_1$, and $\tilde{\vec{v}}_1$ 
represent, respectively, the  pressure, density and fluid velocity in the linear approximation.
Hereafter, a variable with a tilde accent is a dimensionless quantity.
It is convenient to express the excitation in terms of a dimensionless function $\tilde{v}_\text{b}$,
\begin{equation}
        v_\text{ex} = \Mach  c_0 \tilde{v}_\text{b}\,\ee^{-\ii \omega t},\quad \text{with} \quad
        \tilde{v}_\text{b} = \frac{d(\varrho,\varphi)}{d_0}  .
\end{equation}
In acoustofluidic cavities based on piezoelectric ceramics, the peak displacement at the bottom is generally $d_0\approx\SI{1}{\nano\meter}$ in the low-megahertz range~\citep{Tran-Huu-Hue2000}.
Hence, the expected Mach number for a water-filled cavity at $\SI{1}{\mega\hertz}$ is $\Mach\sim 10^{-3}$.

\subsection{Linear acoustics equations}
%\subsection{Governing equations}

A viscous fluid supports both compressional and shear flows.
So we introduce a decomposition
of the fluid velocity as a compressional part ($\tilde{\vec{v}}^\text{c}$) plus a shear part ($\tilde{\vec{v}}^\text{s}$),
\begin{equation}
    \tilde{\vec{v}}_1 
    = \tilde{\vec{v}}_\text{c} + \tilde{\vec{v}}_\text{s}.
    \label{velocity_decomp}
\end{equation}
The compressional component is irrotational, while the shear part is incompressible,
\begin{equation}
        \grad \times \tilde{\vec{v}}_\text{c} = \vec{0} \quad\text{and}\quad 
        \grad \cdot \tilde{\vec{v}}_\text{s} =0.
\end{equation}
Replacing \eqref{densityMach}, \eqref{v1Mach}, and 
\eqref{velocity_decomp} into \eqref{EqDensity}, while
 expressing the dimensionless mass density with \eqref{state_equation} as $\tilde{\rho}_1=\tilde{p}_1$, we obtain the first-order dimensionless pressure as
\begin{equation}
    \label{pressure}
         \tilde{p}_1 = - \frac{\ii c_0}{\omega} \grad \cdot \tilde{\vec{v}}_\text{c}.
\end{equation}
Inserting \eqref{v1Mach} with \eqref{velocity_decomp} into \eqref{EqVelocity} and using \eqref{pressure} followed by the vector operator identity $\grad(\grad\cdot\square) = \nabla^2\square + \grad\times\grad\times\square$,
we arrive at the compressional and shear velocities, respectively,
\begin{subequations}
    \label{FOequations}
    \begin{align}
        \label{v1c}
         \tilde{\vec{v}}_\text{c} & = -
        \frac{\ii c_0 }{\omega} \left(1- \ii \Gamma_\text{c}\right)
        \grad \tilde{p}_1,\\
         \tilde{\vec{v}}_\text{s} &= \frac{\ii 
       \eta_0 }{\omega \rho_0}\nabla^2 \tilde{\vec{v}}_\text{s}. 
       \label{vel_s1}
    \end{align}
\end{subequations}
The compressional damping factor (in the fluid bulk)
is given by
\begin{equation}
\label{gamma_0}
    \Gamma_\text{c} =\kappa_0 \omega \left(\zeta_0 + \frac{4}{3}\eta_0 \right).
\end{equation}
Taking the divergence of  \eqref{v1c} 
and using \eqref{pressure} yields the Helmhotz equation for pressure:
\begin{align}
    \label{Helmoltz_pressure}
        \left(\nabla^2 + k_\text{c}^2\right)\!
        \tilde{p}_1 = 0, \quad \text{with} \quad
        k_\text{c} = \left(
        1 + \frac{\ii}{2}\Gamma_\text{c}\right)k.
\end{align}
The real part of the compressional wavenumber is $k=\omega/c_0$, with the associated acoustic wavelength being $\lambda= 2\upi /k$.
Equation~\eqref{vel_s1} can be recast in the form
\begin{align}
    \label{Helmholtz_vs}
    \left(\nabla^2 + k_\text{s}^2\right)\!
    \tilde{\vec{v}}_\text{s} = \vec{0},\quad
    k_\text{s} =\frac{1+\ii}{\delta},
\end{align}
where 
\begin{equation}
\label{delta}
    \delta=\sqrt{\frac{2 \eta_0}{\rho_0 \omega}}.
    % = \sqrt{\frac{\eta_0 \lambda}{\upi \rho_0 c_0}}
\end{equation}
is the thickness of the viscous boundary layer.
According to table~\ref{tab:PhysicalProperties}, this thickness is $\delta=\SI{0.53}{\micro\meter}$ for water at room temperature ($\SI{25}{\degreeCelsius}$) and $\SI{1}{\mega\hertz}$.

\section{Compressional normal modes}

\subsection{Eigenfunctions of the Helmholtz equation
}

The regular eigenfunctions, i.e., independent solutions to \eqref{Helmoltz_pressure} in cylindrical coordinates, for a rigid cavity resonator can be expressed by~\citet[p. 398]{Morse1948}
\begin{equation}
    \label{p1_sol}
    \tilde{p}_1 = J_n\!\left(k_\varrho \varrho\right)  \left[a_1\cos\!\left(k_z z\right) + a_2 \sin\!\left(k_z z\right)\right]
    \left(b_1\ee^{\ii n \varphi} + b_2 \ee^{-\ii n \varphi}\right),
\end{equation}
where $J_n$ represents the $n$th-order Bessel function with $n\in \mathbb{Z}$.
The coefficients $a_i$ and $b_i$ ($i=1,2$)  will be determined from the boundary conditions~\eqref{boundaryconditions}.
Furthermore, from \eqref{Helmoltz_pressure}  we find that the axial ($k_z$) and circular ($k_\varrho$) wavenumbers are related by
\begin{equation}
    \label{kzkr}
    k_z^2 + k_\varrho^2 = (1+\ii \Gamma_\text{c})\, k^2.
\end{equation}

\begin{table}
\centering
\begin{tabular}{p{.3cm} ccccc}
& \multicolumn{5}{c}{ $m$}\\
\cmidrule{2-6}
\(n\) & \(0\) & \(1\) & \(2\) & \(3\) & \(4\)\\
\cmidrule{2-6}
0 & \multicolumn{1}{r}{\num{0.0000}} & \multicolumn{1}{r}{\num{3.8317}} & \multicolumn{1}{r}{\num{7.0156}} & \multicolumn{1}{r}{\num{10.1735}} & \multicolumn{1}{r}{\num{13.3237}} \\
1 & \multicolumn{1}{r}{\num{1.8412}} & \multicolumn{1}{r}{\num{5.3314}} & \multicolumn{1}{r}{\num{8.5363}} & \multicolumn{1}{r}{\num{11.7060}} & \multicolumn{1}{r}{\num{14.8636}} \\
2 & \multicolumn{1}{r}{\num{3.0542}} & \multicolumn{1}{r}{\num{6.7061}} & \multicolumn{1}{r}{\num{9.9695}} & \multicolumn{1}{r}{\num{13.1704}} & \multicolumn{1}{r}{\num{16.3475}}\\
3 & \multicolumn{1}{r}{\num{4.2012}} & \multicolumn{1}{r}{\num{8.0152}} & \multicolumn{1}{r}{\num{14.5858}} & \multicolumn{1}{r}{\num{17.7887}} & \multicolumn{1}{r}{\num{20.9725}}\\
4 & \multicolumn{1}{r}{\num{5.3176}} & \multicolumn{1}{r}{\num{9.2824}} & \multicolumn{1}{r}{\num{12.6819}} & \multicolumn{1}{r}{\num{15.9641}} & \multicolumn{1}{r}{\num{19.1960}} 
\end{tabular}
\caption{Roots of equation $\dd J_n(x)/\dd x =0$ for  $n,m\le 4$.
\label{tab:BesselRoots}}
\end{table}

\subsection{Natural frequency modes}
\label{sec:naturalFreq}
%\subsection{Nonviscous wavenumber at resonance}

Prior to commencing the analysis of a viscous fluid-filled cavity resonator, it is beneficial to present the natural frequency of the system.
For a rigid cavity with boundary surface denoted by $S$, the pressure satisfies the Neumann boundary condition at the walls, 
\begin{equation}
\vec{n}\cdot\grad \tilde{p}_1\bigl|_{\vec{r}\in S } = 0,
\end{equation}
where $\vec{n}$ is the normal unit vector of the surface $S$ pointing outwardly.
In this case, the axial and circular wavenumbers become
\begin{equation}
    \label{kradkz}
        k_{z}^l = \frac{l \upi}{H}, \quad
        k_{\varrho}^{nm} = \frac{j_{nm}'}{R},
\end{equation}
with   $m,l=0, 1, 2, \dots$
The function $j_{nm}'$ represents a root of equation $\dd J_n(x)/\dd x =0$.
Note that $k_\varrho^{nm}$ is the eigenvalue of both natural modes a positive or negative $n$.
Table~\ref{tab:BesselRoots} shows the roots
considering  mode numbers in the interval $0\le n,m\le5$. 
According to \eqref{kzkr}, the compression wavenumber followed by the angular frequency of the natural mode $(n\,m\,l)$ are given by 
\begin{subequations}
    \begin{align}
     k^{nml}_0 &= \sqrt{\left(k_{z}^l\right)^2 + \left(k_{\varrho}^{nm}\right)^2},
   \label{k0nml}\\
    \label{eq:natural_frequency}
   \omega^{nml}_0 &= c_0 k^{nml}_0.
\end{align}
\end{subequations}

Purely radial waves of mode $(0\,m\,0)$, appear along the radial vector $\vec{e}_\varrho$. 
Likewise, waves aligned with the $z$ axis are present in mode $(0\,0\,l)$. 
Waves corresponding to displacement along the azimuthal direction, represented by $\vec{e}_\varphi$, manifest in mode $(n\,0\,0)$.
Finally, modes characterized by $(n\,m\,0)$ are  designated as circular modes.

\subsection{Scaling parameter}
Viscous effects near solid boundaries are significant due to the steep velocity gradients imposed by the no-slip condition.
For acoustofluidic experimental settings in the megahertz range and above, 
the following condition involving 
the boundary layer thickness is assumed to hold:
\begin{equation}
  \delta \ll R, H, \lambda.
\end{equation} 
We define the scaling parameter for the viscous-boundary depth as
\begin{equation}
    \label{epsilon}
    \epsilon \equiv \frac{ 2\upi \delta}{\lambda} = k \delta \ll 1.
\end{equation}
Note that this parameter scales with frequency as $\epsilon \sim \sqrt{\omega}$.
At a frequency of $\SI{1}{\mega\hertz}$, the scaling parameter for water is  $\epsilon=0.0022$, whereas at $\SI{1}{\giga\hertz}$, it increases to $\epsilon=0.069$.
It is further assume that the peak displacement at the bottom is much smaller than the depth of the boundary layer, $d_0\ll \delta$.  

The compressional damping factor $\Gamma_\text{c}$ can also be expressed in terms of $\epsilon$ by using \eqref{gamma_0},
\begin{equation}
    \Gamma_\text{c} = \left(\frac{2}{3} + \frac{\zeta_0}{2\eta_0}\right)\epsilon^2.
    \label{Gammac}
\end{equation}
Again for water at the same conditions, we have $\Gamma_\text{c} = \num{5.06e-6}$.
Equation \eqref{Gammac} reveals that damping in the cavity bulk is generally very small.

%
% \begin{table}
% \centering
% \begin{tabular}{cc}
% Aspect ratio $\left(\frac{R}{H}\right)$ & Dual mode  \\
% \vspace{-.2cm}
% \\
% \num{1.2197} & $ (0  \, 0  \, 1) + (0  \, 1  \, 0)$   \\
% \num{2.2331} & $(0  \, 0  \, 1) +(0  \, 2  \, 0)$   \\
% \num{0.5861} & $(0  \, 0  \, 1) +(1  \, 0  \, 0)$   \\
% \num{1.6971} & $(0  \, 0  \, 1) +(1  \, 1  \, 0)$   \\
% \num{2.7172} & $(0  \, 0  \, 1) +(1  \, 2  \, 0)$   \\
% \num{0.9722} & $(0  \, 0  \, 1) +(2  \, 0  \, 0)$   \\
% \num{2.1346} & $(0  \, 0  \, 1) +(2  \, 1  \, 0)$   \\
% \num{3.1734} & $(0  \, 0  \, 1) +(2  \, 2  \, 0)$  
% \end{tabular}
% \caption{\label{tab:DualMode} 
% The aspect ratio of microcavities with the corresponding dual modes involving the axial mode $(0\,0\,1)$.}
% \end{table}

%

\section{No-slip boundary conditions}

The core of establishing pressure boundary conditions lies in enforcing the no-slip condition on the cavity walls, expressed as:
\begin{equation}
    \label{noslip}
     \tilde{\vec{v}}_\text{c}(\vec{r}) + \tilde{\vec{v}}_\text{s}(\vec{r}) =
    \begin{cases}
        \tilde{v}_\text{b}\vec{e}_z, \quad & \vec{r}\in S_\text{b},\\
        \vec{0}, \quad & 
        \vec{r}\in
        S_\text{t}  \cup  S_\text{w}.
    \end{cases}
\end{equation}
Here, the ultrasonic actuation with velocity $v_\text{b}$
is exclusively applied at the bottom of the resonator cavity.
In appendix~\ref{app:BC}, we derive the boundary conditions for the linear pressure  based on \eqref{noslip}. 
The pressure should adhere to the following conditions
\begin{subequations}
\label{boundaryconditions}
    \begin{align}
         \left[\partial_z
        + \epsilon \frac{1+\ii}{2k}\left(\partial_{z}^2 + k^2\right)\right] \tilde{p}_1\biggr|_{z=0} &= 0,
        \label{z-zero-BC}
        \\
         \left[\partial_z
        - \epsilon \frac{1+\ii}{2k}\left(\partial_{z}^2 + k^2\right)\right] \tilde{p}_1\biggr|_{z=H} &= 0,
        \label{H-BC}
        \\
        \left\{ \partial_\varrho
        - \epsilon \frac{1+\ii}{2k}
       \left[\frac{1}{\varrho}\partial_\varrho (\varrho\partial_\varrho) + k^2\right]
\right\}\tilde{p}_1\biggr|_{\varrho=R} &= 0.
\label{R-BC}
    \end{align}
\end{subequations}
The boundary conditions stated in \eqref{boundaryconditions} can be interpreted as an impedance boundary condition, specifically incorporating a second-order derivative of pressure.
This result is in agreement with that obtained by \citet[eq. 25]{Bach2018} for cavities with arbitrary curved geometry. 

\begin{figure}
    \begin{center}
         \includegraphics[scale=1]{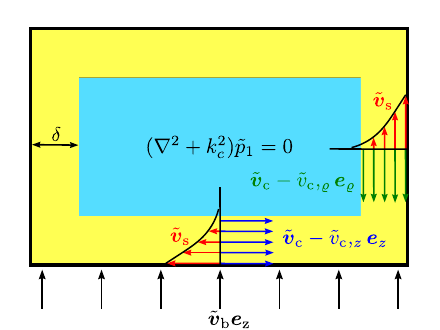} 
    \end{center}
    \caption{The sketch of the acoustic cavity  with rigid walls.
    The yellow area represents the viscous boundary layer, with a depth $\delta$.
    The pressure $\tilde{p}_1$ obeys the Helmholtz equation, shown in light blue. 
    The fluid velocity behaviour within the boundary layer is represented at the bottom $z=0$ and wall $\varrho=R$ through  red (shear velocity) and blue (tangential component of the compressional velocity) arrows. 
    The axial velocity excitation $v_\text{b}\vec{e}_z$ is denoted by black arrows at the bottom.
    \label{fig:resonator2}}
\end{figure}

Figure~\ref{fig:resonator2} presents a schematic representation summarizing the boundary value problem for the pressure field within a cylindrical acoustic cavity. 
A time-harmonic excitation is applied at the cavity's base.

\section{Resonance states}

The acoustic damping  in the boundary layers leads to a reduction in the resonance frequency $\omega^{nml}$.
The shifted frequency can be conveniently expressed as:
\begin{equation}
    \label{k0nml}
    \omega^{nml} = \left(1 - \frac{1}{2}\Gamma^{nml}_\delta\right) \omega_0^{nml},
\end{equation}
where $\Gamma^{nml}_\delta$ is the boundary-layer damping factor. 
Here, we assume $\Gamma_\delta^{nml} \ll 1$ valid for fluids low to moderate viscosity, such as water and vegetable oils.
The associated wavenumber at resonance is $k^{nml} = \omega^{nml}/c_0$. 
As the system operates near a resonance state $(n\,m\,l)$, the axial and circular wavenumber are expected to slightly deviate  from those of the natural  modes:
\begin{subequations}
    \begin{align}
    \label{krappr}
        k_\varrho^{nml} &= k_{\varrho}^{nm} + \Delta {k}_{\varrho}^{nml}, \quad \left|\Delta {k}_{\varrho}^{nml} R \right|\ll 1, \\
        k_z^{nml} &= k_{z}^l + \Delta {k}_{z}^{nml}, \quad \left|\Delta {k}_z^{nml} H \right|\ll 1.
        \label{kznml}
    \end{align}
\end{subequations}
The deviation term in \eqref{krappr} will be determined from the boundary condition \eqref{R-BC}.
The obtained result is then used in~\eqref{kzkr} to derive the axial deviation $\Delta k_z^{nml}$.
%

%\subsection{Circular wavenumber}
%
 In Appendix~\ref{app:radial}, we derive $\Delta k_\varrho^{nml}$  by solving the boundary condition \ref{R-BC}. 
Accordingly, we have
\begin{equation}
   \Delta {k}_{\varrho}^{nml} =
   \begin{cases}
   \pm  \ii \sqrt{\left(1 + \ii\right)\frac{\epsilon^{00l}}{k_0^{00l} R}} k_0^{00l},\quad &n,m=0,\\
   (1+\ii)\frac{\epsilon^{nml}
     }{2k_0^{nml}R}
    \frac{n^2 + (k_{z}^l R)^2 }{n^2-(k_{\varrho}^{nm} R)^2}  k_{\varrho}^{nm}, \quad &\text{otherwise}.
  \end{cases}
  \label{Deltakrnml}
\end{equation}
Here we have introduced the damping parameter $\epsilon^{nml}$ evaluated at the natural frequency $\omega_0^{nml}$:
\begin{subequations}
\begin{align}
\label{epsilon_nml}
\epsilon^{nml} &= k_0^{nml}\delta^{nml}\\
\label{Delta_nml}
\delta^{nml}&=
\sqrt{\frac{2 \eta_0}{\rho_0 \omega_0^{nml}}}, 
\end{align}
\end{subequations}

We now proceed to obtain the axial wavenumber at resonance. 
Using \eqref{kzkr} we find    
\begin{equation}
    \label{kz2}
   \left[k_z^{nml}(k)\right]^2 =
    \left(1+\ii \Gamma_\text{c}^{nml}\right)k^2 - \left(k_\varrho^{nml}\right)^2,
\end{equation}
where the compressional damping factor at the natural frequency values $\Gamma_\text{c}^{nml}\bigl|_{\epsilon=\epsilon^{nml}}$.
We replace  \eqref{krappr} into  \eqref{kz2}, use $k=\omega/c_0$, and expand the result
around the  resonance frequency $\omega=\omega^{nml}$, while keeping the result to its lowest-order,
\begin{align}
\nonumber
    \left[k_z^{nml}(\omega)\right]^2&= \left(k_{z}^l\right)^2 + \frac{2 k_0^{nml}}{c_0}(\omega-\omega^{nml}) + (\ii \Gamma_\text{c}^{nml}
    -\Gamma_\delta^{nml})
    \left(k_0^{nml}\right)^2 -
    2 k_{\varrho}^{nm}
    \Delta k_\varrho^{nml}\\
    &\phantom{=}-\delta_{n,0}\delta_{m,0}
    \left(\Delta k_\varrho^{00l}\right)^2,
    \label{kz3}
\end{align}
where $\delta_{ij}$ is the Kronecker delta function.
Note that $(\Delta k_\varrho^{00l})^2$ varies linearly with $\epsilon_{nml}$.
If the axial mode is suppressed ($l=0$), we have
\begin{align}
\left[k_z^{nm0}(\omega)\right]^2 =
       \frac{2 k_0^{nm0}}{c_0}(\omega-\omega^{nm0}) + \left(\ii
     \Gamma_\text{c}^{nm0}-\Gamma_\delta^{nm0}\right)(k_0^{nm0})^2 -
    2 k_{\varrho}^{nm}\Delta k_\varrho^{nm0}.
    \label{deltakz0}
\end{align}

\section{Single-mode configuration}

\subsection{Vortex excitation}
In accordance with \eqref{v1ex}, the velocity excitation required to generate a mode with positive vorticity in the cavity is described as 
\begin{equation}
\label{vex2}
\tilde{v}_\text{b} = J_{n}\!\left(k_{\varrho}^{nm}\varrho\right) \ee^{ \ii n \varphi}.
\end{equation}
The induced pressure amplitude rotates counterclockwise when observed from above the $xy$ plane.
In this case, the azimuthal coefficients in \eqref{p1_sol} become $b_1=1$ and $b_2=0$.
Conversely, when the angular factor is $\ee^{-\ii n\varphi}$, the induced pressure rotates clockwise from the same viewpoint, and 
$b_1=0$ and $b_2=1$.
For the case where $n=0$, the excitation produces a nonvortex mode within the resonator, with $b_1=b_2=1/2$.

\subsection{Pressure coefficients}
Substituting \eqref{p1_sol} and \eqref{vex2} into \eqref{z-zero-BC} and \eqref{H-BC}, we arrive at a system of linear equations for the pressure coefficients $a_1$ and $a_2$:
\begin{subequations} 
     \label{Axb}
    \begin{align}
    &\mathsfbi{A}
\left( \begin{matrix}
a_1 \\ a_2
\end{matrix}
\right) = \left( \begin{matrix}
  \ii  (k H)^2 \\ 0 
\end{matrix}
\right),\\
        &\mathsfbi{A} =
        \left( 
        \begin{matrix}
        \epsilon  \chi &  k_z k H^2 \\  
          k_z k H^2 \sin(k_z H) + \epsilon \chi  \cos (k_z H)&  
          - k k_z H^2 \cos (k_z H)+ \epsilon \chi 
        \sin( k_z H)
        \end{matrix}
        \right),
    \end{align}
\end{subequations}
with
$
\chi = (1+\ii) \left(k^2-k_z^2\right) H^2/2$.
Terms of order $\epsilon\Gamma_\text{c}$ were discarded.
Also, we have taken the approximation  $J_n(k_\varrho^{nml} \varrho)\approx J_n(k_{\varrho}^{nm} \varrho)$.
According to the Crammer's rule, the solution of  \eqref{Axb} is given by
\begin{subequations}
    \begin{align}
    \label{a1General}
        a_1 &= \frac{\ii k}{k_z} \frac{ k k_z H^2 \cos (k_z H) - \epsilon \chi  \sin (k_z H)}
     { k k_z H^2 \sin (k_z H) + 2 \epsilon \chi  \cos (k_z H)},\\
     a_2 &= \frac{\ii k}{k_z} \frac{k k_z H^2 \sin (k_z H) + \epsilon\chi  \cos (k_z H)}
     {k k_z H^2 \sin (k_z H) + 2 \epsilon \chi  \cos (k_z H)}.
    \end{align}
\end{subequations}
It is important to compare the magnitude of the pressure coefficients. 
Referring to \eqref{kznml}, we observe that $\cos(k_z z) \approx \cos(k_z^l z)$ and $\sin(k_z z) \approx \sin(k_z^l z)$. 
This implies that $a_2/a_1 \sim \epsilon$, allowing us to safely neglect the term $a_2 \sin(k_z z)$ in \eqref{p1_sol}.
Therefore, the pressure of near the resonance is assumed to be
\begin{equation}
      \tilde{p}_1^{nml}(\vec{r}) = a_1 J_n\!\left(k_\varrho^{nm} \varrho\right)
      \cos\!\left(k_z^l z\right) 
      \ee^{\ii n \varphi}.
      \label{PressureCoeff}
\end{equation}

The amplitude coefficient $a_1$ and the boundary-layer factor $\Gamma_\delta^{nml}$ are determined
in appendix~\ref{app:PressureCoefficients}.
Gathering the result, we have
\begin{subequations}
\begin{align}
    &a_1 =   \frac{\ii (2-\delta_{l,0})}{k_0^{nml}H} 
    \frac{\tfrac{1}{2}\omega^{nml}}{\omega - \omega^{n m l} + \tfrac{1}{2} \ii\,\omega^{nml} \Gamma^{n m l} },
    \label{aznml2}\\
    \label{GammaTotal}
    &\Gamma^{nml} = \Gamma_\delta^{nml} + \Gamma_\text{c}^{nml},\\
     &\Gamma_\delta^{nml} =
     \epsilon^{nml}
     \left\{
     \left[\frac{2-\delta_{l,0}}{k_0^{nml} H}  + 
    \frac{1}{k_0^{nml} R}
    \frac{(k_{z}^l R)^2 + n^2 }{(k_{\varrho}^{nm} R)^2-n^2} 
     \right] \left(\frac{k_{\varrho}^{nm}}{k_0^{nml}}\right)^2
     + \frac{ \delta_{n,0} \delta_{m,0}}{ k_0^{nml} R}
  \left(\frac{k_{z}^{l}}{k_0^{nml}}\right)^2
     \right\}.
     \label{Gammanml2}
\end{align}
\end{subequations} 
The total damping factor $\Gamma^{nml}$ is determined by using the wavenumbers related to 
the natural frequency $\omega_0^{nml}$.

% As $\Gamma_\delta^{nml} \sim \epsilon^{nml}$
% and 
% $\Gamma_\text{c}^{nml} \sim (\epsilon^{nml})^2$, 
% the damping within the boundary layer is in general more prominent than the bulk attenuation,
% $\Gamma_\delta^{nml} \gg \Gamma_\text{c}^{nml}$.

% Using \eqref{epsilon_nml}, $k_0^{00l}=l \pi/H$,  and $k_0^{nm0}=j_{nm}'/R$, we find the boundary-layer damping factor for a purely axial and vortex-radial mode as
% \begin{subequations}
% \label{Gammas}
%     \begin{align}
%     \label{Gamma_00l}
%         \Gamma_\delta^{l} &=\Gamma_\delta^{00l} = \frac{\delta^{l}}{R}, \quad l> 0,\\
%         \label{Gamma_nm0}
%                 \Gamma_\delta^{nm} &=\Gamma_\delta^{nm0} = f^{nm}
%    \frac{\delta^{nm}}{H},
%          \\
%    f^{nm} &= 
%     1+ \frac{H}{R}\frac{n^2}{(j_{nm}')^2-n^2}
%     \end{align}
% \end{subequations}
% with 
% $\delta^l=\delta^{00l}$ and
% $\delta^{nm}=\delta^{nm0}$.

\subsection{Pressure field}
We now replace  \eqref{aznml2} into  \eqref{PressureCoeff} to obtain the pressure amplitude of the single-vortex mode $(n\,m\,l)$,
\begin{subequations}
    \label{pres}
    \begin{align}
        {p}_1^{nml} 
        &=  {A}^{nml}  R^{nml}
        \Psi^{nml},
        \label{p1lm}
        \\
        \label{Anml}
         {A}^{nml} &=
         \frac{2-\delta_{l,0}}{\Gamma^{nml} k_0^{nml}H},\\
        {R}^{nml} &=\frac{1}{2}\frac{
        \omega^{nml}\Gamma^{nml} }
        { 
        \omega - \omega^{nml}\left(1
        + \frac{1}{2}\ii \,
        \Gamma^{nml}\right)},\\
        \Psi^{nml} &= J_n\!\!\left(k_\varrho^{nm} \varrho\right) 
        \cos\!\left(k_z^l z\right) \ee^{\ii \left(n \varphi+\frac{\pi}{2}\right)}.
        \label{Psi}
    \end{align}
\end{subequations}
where 
${A}^{nml}$ is the pressure amplitude,
${R}^{nml}$ is the resonance function
and $\Psi^{nml}$ is the cavity's wavefunction.
The  pressure amplitude scales inversely with the total damping factor $(\Gamma^{nml})^{-1}$, and the height-to-wavelength ratio $(k_0^{nml} H)$.
At the resonance frequency, we have
$    R^{nml}|_{\omega=\omega^{nml}} = -\ii$. Hence the pressure becomes
\begin{equation}
p_1^{nml}|_{\omega=\omega^{nml}}= -\ii {A}^{nml} 
        \Psi^{nml}
\end{equation}
We note also that the pressure phase  is composed of three components, namely, the  resonance   $(\varphi_\text{r}^{nml})$, spatial ($\varphi_\text{s}^{nml}$), and
 vortex  $(n\varphi)$ phase:
\begin{equation}
    \varphi^{nml} = \varphi_\text{r}^{nml} + 
    \varphi_\text{s}^{nml}+
    n\varphi + \frac{\pi}{2}.
\end{equation}

\subsection{Fluid velocity fields}
We can determine the fluid velocity in the fluid bulk by using  \eqref{v1c},
\begin{equation}
\tilde{\vec{v}}_\text{c}^{nml}   =
     -\ii {A}^{nml} {R}^{nml}\,
    \tilde{\grad}\Psi^{nml},
\label{v1cnml}
\end{equation}
where $\tilde{\grad}=(k_0^{nml})^{-1}\grad$ is the scaled gradient to the natural wavenumber.
Simultaneously,
the shear velocity  near the boundaries can be calculated by applying 
 \eqref{v1cnml} in combination with  \eqref{shear_decomp}
and \eqref{shear_velocities},
\begin{equation}
    \tilde{\vec{v}}_\text{s}^{nml}
        = -\ii {A}^{nml} {R}^{nml}
        \biggl[
        \left(
            \ee^{\ii k_\text{s} z} 
            + (-1)^l \ee^{\ii k_\text{s} (H-z)} \right)
            \tilde{\grad}_{\!\vec{e}_z}
            + \ee^{\ii k_\text{s} (R-\varrho)} 
     \tilde{\grad}_{\!\vec{e}_\varrho}
            \biggr]\Psi^{nml},
\end{equation}
with $\tilde{\grad}_{\!\vec{n}}= \left(\mathsfbi{I} - \vec{n}\vec{n}\right)\cdot\tilde{\grad}$
being the scaled surface gradient.
The shear velocity is confined in the viscous boundary as it decays exponentially away from the cavity walls.

\section{Dual-mode configuration}

Thus far, our focus has been on single-mode resonances.
Nevertheless, it is possible to use a dual mode excitation inside the cavity.
Consider the natural frequency \(\omega_0^{nml}\) associated with the single mode \((n\,m\,l)\),
which is excited through the applied vibration at the resonator's bottom as described by \eqref{vex2}.
This natural frequency can also be linked to two other  modes \((n_1\,m_1\,l_1)\) and \((-n_1\,m_1\,l_1)\) by
the relation $\omega_0^{nml} = \omega_0^{|n_1|ml}$.
% In such a way that modes with \(\pm n_1\) are degenerate because the circular wavenumber is an even function of \(n\), e.g., \(k_\varrho^{(-n)m} = k_\varrho^{nm}\).
% Moreover, according to \eqref{Gammanml2} the cavity damping factor also satisfies
% $\Gamma^{(-n) m l} = \Gamma^{n m l}$.
% %
Two possible dual-mode configurations arise, which are denoted, respectively by
$    \left( n\,m\,l \right)+\left(n_1\,m_1\,l_1 \right)$ and
$ \left( n\,m\,l \right)+\tfrac{1}{2}\left[\left(n_1\,m_1\,l_1 \right)+\left(-n_1\,m_1\,l_1 \right)\right]$.
For the sake of brevity, we will analyse the simplest dual-mode configuration corresponding to the mode $    \left( n\,m\,l \right)+\left(n_1\,m_1\,l_1 \right)$.

When the single-mode natural frequencies are equal, we  have the wavenumber relation
\begin{equation}
    k_0^{nml} = k_0^{n_1m_1l_1}.
    \label{dual-freqWavenumber}
\end{equation}
Using \eqref{kradkz}, we find that this configuration can only occur when cavity's aspect ratio satisfies the condition
\begin{equation}
\label{AR_dualmode}
\frac{R}{H} = \frac{1}{\upi}\sqrt{\frac{(j_{n_1 m_1}')^2 - (j_{n m}')^2}{l^2 - l_1^2}}.
\end{equation}
Using this relation we can induce the dual-mode configuration in a cavity by changing excitation velocity in \eqref{vex2} to
\begin{equation}
\label{vex3}
\tilde{v}_\text{b} = J_{n}\!\left(k_{\varrho}^{nm}\varrho\right) \ee^{ \ii n \varphi} + 
J_{n_1}\!\left(k_{\varrho}^{n_1m_1}\varrho\right) \ee^{ \ii n_1 \varphi}.
\end{equation}
Note that the actual resonance frequencies of each single mode are approximately equal, $\omega^{nml}\approx \omega^{n_1m_1l_1}$.
The pressure inside the cavity given by
\begin{equation}
    \label{pn1n2}
        \tilde{p}_1=\tilde{p}_{n ml}^{n_1 m_1l_1} =\tilde{p}_1^{n ml} +
         \tilde{p}_1^{n_1 m_1l_1}.
\end{equation}
At this point, we proceed with our analysis at the resonance frequency $\omega=\omega^{nml}$.
From \eqref{p1lm}, we can express \eqref{pn1n2} by
\begin{equation}
    \label{DualPressure}
    \tilde{p}_{n ml}^{  n_1 m_1 l_1} = 
        A^{n m l} R^{n m l}  \Psi^{n m l} +
        {A}^{n_1m_1 l_1}R^{n_1 m_1 l_1}  \Psi^{ n_1 m_1l_1}.
\end{equation}

% When $n_1=0$, it is observed that the standing wave pressure becomes  $\tilde{p}_1^{\left\langle n_1m_1l_1\right\rangle}=0$.
% Consequently,  
% \begin{equation}
%     \tilde{p}_{n ml}^{\langle 0m_1l_1\rangle} = \tilde{p}_1^{n ml}.
%     \label{DualModeN1Zero}
% \end{equation}

% A nonvortex excitation, described by $v_\text{b}^{0m} = J_0(k_\varrho^{0m}\varrho)$, can produce a dual-mode configuration as the aspect ratio is 
% \begin{equation}
% \frac{R}{H} = \frac{j_{0m}'}{l \upi}.
% \end{equation}
% This configuration produces the dual-mode pressure:
% \begin{equation}
%       {p}_{0 m l}^{\langle n_1 m_1 0\rangle}(\vec{r},\omega)
%         =  A^{0 m l} R^{0 m l}(\omega)\, \Psi^{0 m l}(\vec{r}) +
%         {A}^{n_1 m_1 0}  R^{n_1 m_1 0}(\omega)\, \Psi^{\langle n_1 m_1 0\rangle}(\vec{r}).
% \end{equation}
% We see that the nonvortex excitation yields circular standing wave. 

% Hence, the total fluid velocity of a normal mode is
% \begin{equation}
%     v_1^{nml} = \left(\tilde{\vec{v}}_1^\text{c}\right)^{nml}
%     + \left(\tilde{\vec{v}}_1^\text{s}\right)^{nml}.
% \end{equation}

% Again, a dual-mode excitation in the microcavity yields the fluid velocity as
% \begin{equation}
%     v_1^{nml+n_2m_2l_2} = v_1^{nml} + v_1^{n_2m_2l_2}.
% \end{equation}

\section{Nonlinear acoustic phenomena}

We present important phenomena governed by acoustic fields of the magnitude $\bigO(\Mach^2)$, which include considerations of energy density,
radiation force, and spin angular momentum.
These effects are described by time-averaged quantities derived from the product of first-order pressure and velocity fields. 
It can be demonstrated that the time-average of the product of two linear complex-amplitude fields \(f(\vec{r})\,e^{-\ii \omega t}\) and \(g(\vec{r})\,e^{-\ii \omega t}\) is given by
\begin{equation}
\overline{f(\vec{r})\,\ee^{-\ii \omega t} g(\vec{r})\,\ee^{-\ii \omega t}} =
\frac{\omega}{2\upi} 
\int_{-\tfrac{\upi}{\omega}}^{\tfrac{\upi}{\omega}} 
\re\!\left[f(\vec{r})\,\ee^{-\ii \omega t}\right]
\re\!\left[g(\vec{r})\,\ee^{-\ii \omega t}\right] \dd t
=
\frac{1}{2} \re\!\left[f(\vec{r}) g^*(\vec{r})\right],
\end{equation}
where the asterisk $^*$ denotes complex conjugation.
To facilitate the upcoming analysis, the second-order fields will be expressed in terms of the  pressure amplitude squared $|\tilde{p}_1|^2$.
Moreover, given the disparity in dimensions between the boundary layer thickness and the microcavity size, the predominant velocity component of second-order acoustic fields analyzed here arises from the compressional fluid velocity.

\subsection{Acoustic energy density}
The time-averaged acoustic energy density, often referred to as the energy landscape, offers a spatial representation of energy distribution within the cavity. 
This provides valuable insights into the system's behavior and performance.
The energy landscape ($E$) arises from the  contributions of both kinetic ($E_\text{k}$) and potential ($E_\text{p}$)  acoustic energy densities,
\begin{subequations}
    \label{EnergyDensity}
    \begin{align}
        {E} &= {E}_\text{k} +{E}_\text{p},\\
        \label{KineticEnergyDensity}
        {E}_\text{k}&= E_0\left|\tilde{\vec{v}}_\text{c}\right|^2,\\
        {E}_\text{p}&=E_0 \left|\tilde{p}_1\right|^2,
        \label{PotentialEnergyDensity}
    \end{align}
\end{subequations}
where $E_0 = \Mach^2 \rho_0 c_0^2/4$ is the characteristic energy density of the cavity.
%The estimated characteristic energy water-filled cavity at $\SI{1}{\mega\hertz}$ is $E_0=\SI{220.4}{\joule\per\meter\cubed}$.
%
Using \eqref{v1c} allows us to express the kinetic energy density as
\begin{equation}
    {E}_\text{k}= 
    \frac{E_0}{k^2}  \grad\! \tilde{p}_1 \cdot  \grad\! \tilde{p}_1^*=
    \frac{E_0}{k^2}\left(\frac{1}{2}\nabla^2|\tilde{p}_1|^2-\re[\tilde{p}_1^*\nabla^2 \tilde{p}_1]\right)=
    E_0\left(\frac{1}{2k^2}\nabla^2 + 1\right)
    |\tilde{p}_1|^2.
    \label{KineticEnergyDensity2}
\end{equation}
Here we have used $\nabla^2\tilde{p}_1=-k^2 \tilde{p}_1$.
Hence, the  energy density reads
    \begin{align}
    \label{EnergyDensity2}
        {E} = {E}_\text{k} + {E}_\text{p} = 
            2 E_0 \left(\frac{1}{4k^2}\nabla^2 + 1\right) |\tilde{p}_1|^2.
    \end{align}
We define the dimensionless energy density as $\tilde{E}=E/E_0$.
Using \eqref{p1lm} and \eqref{DualPressure}, we obtain the single-mode energy density   as
\begin{equation}
    \label{Enml_energy}
        \tilde{E}^{nml} = 
        2\left|A^{nml}R^{nml} \right|^2
\left(\frac{1}{4}\tilde{\nabla}^2 + 1\right)
\left|\Psi^{nml}\right|^2,
\end{equation}
where $\tilde{\nabla}^2=k^{-2}\nabla^2$ is the scaled Laplacian.
Inserting \eqref{DualPressure} into \eqref{EnergyDensity2},
we obtain
the energy density of the dual mode as
\begin{equation}
\tilde{E}^{n_1 m_1l_1}_{nml} =
\tilde{E}^{nml} + \tilde{E}^{ n_1 m_1l_1} +
\tilde{E}_\text{cross}.
\label{EDualMode}
\end{equation}
The cross term of the energy density is given by
\begin{equation}
\tilde{E}_\text{cross}=
4 A^{nml} A^{n_1m_1 l_1}
\re \! \left[ \left(R^{nml}\right)^* R^{n_1m_1 l_1}
 \left(\frac{1}{4}\tilde{\nabla}^2 + 1\right)
\left({\Psi^{nml}}\right)^*
\Psi^{n_1 m_1l_1}\right].
\label{crossTermEnergy}
\end{equation}
While the single-mode energy density does not depend on the azimuthal angle $\varphi$, the crossed energy does it.

\subsection{Dissipated power}

The time-averaged dissipated power $P^{nml}$ for a specific mode  $(n\,m\,l)$ is linked with the energy stored in the cavity $(E_\text{stored})$ through the concept of the cavity's quality factor, or $Q$ factor. 
As discussed by \citep{Hahn2015}, this relationship is  expressed as
\begin{equation}
    Q = 
    \frac{\omega^{nml}}{P^{nml}} \int_{V_0} E^{nml} \, \dd V
    =
    \frac{\omega^{n m l}}{\Delta \omega^{n m l}}.
    %=
    %\frac{1}{\Gamma^{n m l} },
    \label{Qnm}
\end{equation}
Here, the stored energy is derived from the  integration of the energy density over the cavity volume  $V_0$.
The term $\Delta \omega^{n ml}$ represents the $\SI{-3}{\decibel}$ bandwidth of the resonance line, which is determined by solving the equation $|R^{n m l}(\omega)|^2 = \tfrac{1}{2}$, subsequently leading to $\Delta \omega^{n m l} = \omega^{n m l} \Gamma^{n m l}.$
Replacing the bandwidth  into  \eqref{Qnm} gives
the $Q$ factor as
\begin{equation}
    Q =\frac{1}{\Gamma^{nml}}.
\end{equation}
Consequently, the dissipated power  by the cavity is
\begin{equation}
    P^{nml}= \Gamma^{nml}\omega^{nml}\int_{V_0} E^{nml} \, \dd V. 
\end{equation}

\subsection{Acoustic radiation force}
An important application of acoustofluidic microcavities is the contactless manipulations of small particles.
We consider a spherical particle of volume $V_\text{p}$, isentropic compressibility $\kappa_\text{p}$, and density $\rho_\text{p}$.
Moreover,
the particle in consideration is assumed to be much smaller than the wavelength.
In this case, the acoustic radiation force exerted on a particle at position $\vec{r}$  is given by~\citep{Gorkov1962}
\begin{subequations}
    \begin{align}
    \label{RadiationForce}
    \vec{F}&= -V_\text{p}\grad U,\\
    U &=  f_\kappa {E}_\text{p}
    -\frac{3}{2}f_\rho {E}_\text{k},
    \label{Unml}\\
    f_\kappa &= 
    1 - \frac{\kappa_\text{p}}{\kappa_0}, \quad
    f_\rho  = 2 \frac{\rho_\text{p} - \rho_0}{2\rho_\text{p}+ \rho_0},
    \end{align}
\end{subequations}
where $U$ is the  Gor'kov potential density, and $f_\kappa$ and $f_\rho$ are, respectively, the compressional and density contrast factors.
After inserting the  potential \eqref{PotentialEnergyDensity} and kinetic \eqref{KineticEnergyDensity} energy densities into \eqref{Unml}, we find the dimensionless Gor'kov potential density $(\tilde{U}=U/E_0)$  as
\begin{equation}
    \tilde{U} =  
    \left[ f_\kappa 
        - \frac{3}{2} f_\rho\left(
        \frac{1 }{2 }
        \tilde{\nabla}^2 +1 \right)\right]
        \left|\tilde{p}_1\right|^2.
        \label{Gorkov_p1}
\end{equation}
Substituting \eqref{p1lm} into
this equation yields the single-mode Gor'kov potential 
\begin{equation}
\label{Unml}
    \tilde{U}^{nml} =  
\left|A^{nml} R^{nml} \right|^2
    \left[ f_\kappa 
        - \frac{3}{2} f_\rho\left(\frac{1 }{2 }
        \tilde{\nabla}^2+1
        \right)\right]
        \left|\Psi^{nml}\right|^2.
\end{equation}

The Gor'kov potential of a dual mode  is obtained by replacing \eqref{pn1n2} into  \eqref{Gorkov_p1},
\begin{equation}
   \tilde{U}^{n_1 m_1l_1}_{nml} =
\tilde{U}^{nml} + \tilde{U}^{ n_1 m_1 l_1} +
\tilde{U}_{\text{cross}}.
\label{UDualMode}
\end{equation}
The cross term of the dual-mode potential
is
\begin{align}
\tilde{U}_{\text{cross}} =
2  A^{nml} A^{n_1m_1 l_1} \re \! \left\{\left(R^{nml}\right)^* R^{n_1m_1 l_1}
\left[ f_\kappa 
        - \frac{3}{2} f_\rho\left(
        \frac{1 }{2 }
        \tilde{\nabla}^2 +1 \right)\right]
{\Psi^{nml}}^*
\Psi^{n_1 m_1l_1}\right\}.
\label{crossTermU}
\end{align}

\subsection{Spin angular momentum}
% The spin angular momentum manifests uniquely across different domains. 
% %
% In classical electromagnetic waves and photons, this spin arises from the circular polarization of electric and magnetic fields [1]. 
% %
% Electron spin, on the other hand, can be attributed to a circulating energy flow within the Dirac wave field [2]. Recently, the concept of spin has been extended to acoustic beams, where it is postulated and experimentally validated as resulting from the circulation of fluid velocity [3]. 
% %
% Subsequently, both spin and orbital angular momenta have been subject to theoretical scrutiny within monochromatic acoustic wave fields in a uniform medium [4]. 
% %
% Prior to these investigations, it was observed that longitudinal spin, characterized by rotation aligned with the propagation direction within an acoustic Bessel beam, could induce an acoustic radiation torque on a subwavelength absorbing spherical particle [5].

The spin angular momentum density of an acoustic wave can be expressed by~\citep{Bliokh2019,Lopes2020}
\begin{equation}
    \vec{S} = \frac{\rho_0}{2 \omega}\im\!\left[{\vec{v}}_\text{c}^* \times  {\vec{v}}_\text{c}\right],
%    \frac{2E_0}{\omega} \im\!\left[\tilde{\vec{v}}_\text{c}^* \times  \tilde{\vec{v}}_\text{c}\right],
\end{equation}
which is given in the SI  units of  $\si{\joule\second\per\meter\cubed}$.
Noting from \eqref{v1c} that ${\vec{v}}_\text{c}\approx -\tfrac{1}{\ii \rho_0 \omega}\grad\! {p}_1$.
We can write the acoustic spin as
\begin{equation}
\vec{{S}} =\frac{1}{2\rho_0\omega^3}\Imag \!\left[ \grad\! {p}_1^* \times \grad\! \tilde{p}_1\right]=
\frac{1}{2\rho_0\omega^3}\Imag \!\left[\grad \times \left({p}_1^*\grad \!{p}_1\right) \right]=
\frac{1}{\omega^2}\grad \times \tfrac{1}{2}\Real\! \left[ {p}_1^* {\vec{v}}_c\right]=
\frac{1}{\omega^2}\grad \times \vec{I}.
\label{Spin2}
\end{equation}
It is important to note that the spin angular momentum is a measure of the circulation of acoustic intensity, defined as $\vec{I} = \tfrac{1}{2} \text{Re} [ {p}_1^* \vec{v}_c ]$. The acoustic spin density forms a solenoidal field, meaning $\nabla \cdot \vec{S} = 0$. When a particle is present in the wave path, it modifies the spin angular momentum, leading to the generation of a resultant acoustic torque on the particle, known as spin-torque. \cite{Silva2014} established a direct relationship between the radiation torque and the acoustic spin density $\vec{S}$. This spin-torque induces rotational motion in the particle, causing it to rotate in a plane perpendicular to the direction of the acoustic spin.

We define the dimensionless acoustic spin density as $\tilde{\vec{S}}= \omega \vec{S}/E_0$.
To obtain the spin of a single vortex mode at the resonance, $\omega=\omega^{nml}$, we substitute \eqref{p1lm}  into \eqref{Spin2},
\begin{equation}
\label{SingleSpin}
       \tilde{\vec{S}}^{nml}
    = 2 
     \left|{A}^{nml}R^{nml} \right|^2\Imag\! \left[
    \tilde{\grad} \times \left(
{\Psi^{nml} }^*
\tilde{\grad}\Psi^{nml}
\right)
    \right].
\end{equation}
%We observe that modes lacking vorticity $(n=0)$ do not possess spin, $\vec{S}^{0ml}=\vec{0}$. 
%
For the dual mode configuration one has
\begin{align}
    \tilde{\vec{S}}_{nml}^{n_1m_1l_1}
    = \tilde{\vec{S}}^{nml} + \tilde{\vec{S}}^{ n_1 m_1l_1} +
    \tilde{\vec{S}}_{\text{cross}},
\label{SDualMode}
\end{align}
with the spin cross-term being
\begin{align}
\nonumber
\tilde{\vec{S}}_\text{cross} &=
2 A^{nml} A^{n_1m_1l_1}
    \Imag\biggl\{
    \tilde{\grad} \times 
    \biggl[ \left(R^{nml}\right)^* R^{n_1m_1 l_1}
\left(\Psi^{nml} \right)^*
\tilde{\grad}\Psi^{n_1m_1l_1} \\
&+ \left(R^{n_1m_1l_1}\right)^* R^{nm l}\left(\Psi^{n_1 m_1l_1}\right)^* \tilde{\grad}{\Psi^{nml} }
\biggr]\biggr\}.
\label{crossTermS}
\end{align}

\section{Acoustofluidic cylindrical resonators}

The utilization of acoustofluidic cylindrical resonators stands as a pivotal technique in processes such as cell enrichment and precise particle patterning. 
These resonators serve to adeptly form particle aggregates within a defined levitation plane.
The number of levitation planes equals are determined by the axial mode number $l$.

\subsection{Acoustic single modes}
Two modes of particular interest are the  axial and the transverse mode.
The  energy landscape of these modes at the resonance $(\omega=\omega^{nml})$   are derived by substituting  \eqref{pres} into \eqref{Enml_energy}.
The result yields, respectively,
\begin{subequations}
    \begin{align}
    \label{EnergyDensity_l}
        \tilde{E}^{l} &=
        \left|A^l R^{l} \right|^2,
        \\
        \label{EnergyDensity_nm0}
        \tilde{E}^{nm} &=   
        \left|A^{nm} R^{nm} \right|^2  
    \left[J_n\!\left(\tilde{\varrho}^{nm} \right)\left[\frac{J_n'\!\left(\tilde{\varrho}^{nm}\right)}{\tilde{\varrho}^{nm}} +
    J_n''\!\left(\tilde{\varrho}^{nm}\right)\right]
    +\left[J_n'\!\left(\tilde{\varrho}^{nm}\right)\right]^2  + 2 J_n^2\!\left(\tilde{\varrho}^{nm}\right)
    \right],
    \end{align}
\end{subequations}
where $A^l=A^{00l}$, $A^{nm}=A^{nm0}$, $R^{l}=R^{00l}$, and $R^{nm}=R^{nm0}$, and  $\tilde{\varrho}^{nm}=k_\varrho^{nm}\varrho$.
Note that  the energy density of the vortex mode exhibits circular symmetry, while it remains spatially constant for the axial mode.

Now we turn our analysis to  the Gor'kov potential near the resonance frequency $\omega\approx \omega^{nml}$.
For the axial mode $(0\,0\,l)$, this potential 
is obtained from \eqref{Unml}.
Accordingly, we have
\begin{equation}
        \label{U001}
        \tilde{U}^{l} = \frac{1}{2} \left|A^l R^l\right|^2
                \left[\left(2f_\kappa + 3f_\rho\right) \cos^2 \!
                \tilde{z}^l
                - 3f_\rho \right],
\end{equation}
where $\tilde{z}^l=k_z^l z$.
% %
The particles will be trapped at the Gor'kov potential minimum by the action of the acoustic radiation force.
When $2f_\kappa + 3f_\rho>0$, the potential minima is in the axial pressure nodes at
\(z_\text{min}=(2\ell+1)H/2l\), with $\ell=0,1,\dots,l-1$.
If $2f_\kappa + 3f_\rho<0$, the minima is at the pressure antinodes in $z_\text{min} =  \ell H/l$, where $\ell=0,1,\dots,l$. 
The  axial radiation force per unit length is obtained by taking minus the gradient of  \eqref{U001}:
\begin{equation}
\label{Fzrad}
    {F}_z^l =\frac{E_0 V_\text{p}}{2} \left|A^l R^l\right|^2
                \left(2f_\kappa + 3f_\rho\right) \sin\!\left( 2\tilde{z}^l \right).
\end{equation}

Considering the transverse mode $(n\,m\,0)$,
we find from \eqref{Unml} that the Gor'kov potential is
    \begin{align}
    \nonumber
        \tilde{U}^{nm} &= \frac{1}{2}
        \left|A^{nm} R^{nm} \right|^2
        \biggl\{
        \left(2f_\kappa - 3f_\rho\right) 
\left[J_n\!\left(\tilde{\varrho}^{nm}\right)\right]^2
        \\
         \label{Unm0}
        &
        - 3 f_\rho 
\left[J_n'\!\left(\tilde{\varrho}^{nm}\right)\right]^2 - 3 f_\rho 
J_n\!\left(\tilde{\varrho}^{nm} \right)        
\biggl[ 
J_n''\!\left(\tilde{\varrho}^{nm} \right)
+ \frac{J_n'\!\left(\tilde{\varrho}^{nm}\right)}{\tilde{\varrho}^{nm}} 
\biggr]\biggr\}.
    \end{align}
The Gor'kov potential here has no azimuthal dependence on $\varphi$.
For the sake of brevity, we  will analyse the radial trapping around the symmetry axis of the cavity only.
To do so, we use the asymptotic form of the Bessel function for small arguments, $J_n(x) \approx 2^{-n} x^n ( 1/n! - 1/4[(n+1)!])$, with $x \rightarrow 0$.
The radial radiation force of the nonvortex and first vortex mode ($n=0,1$) are given by 
\begin{subequations}
    \begin{align}
        {F}_\varrho^{0m} &= f_\kappa
       E_0  V_\text{p}\left|A^{0m} R^{0m} \right|^2    k_\varrho^{0m}  \varrho
        + \bigO\!\left(\varrho^3\right)
        \\
         {F}^{1m}_\varrho &= -\left( 2f_\kappa + 3f_\rho\right) E_0 V_\text{p} 
         \left|A^{1m} R^{1m} \right|^2 
         \varrho + \bigO\!\left(\varrho^3\right),  
    \end{align}
\end{subequations}
Some conclusions can be drawn from the analysis here.
Particles in the non-vortex mode $(0\,m)$ can either be repelled from $(f_\kappa > 0)$ or attracted to $(f_\kappa < 0)$ the cavity axis. In the first-vortex mode $(1\,m)$, particles will be trapped at the cavity axis if $2f_\kappa + 3f_\rho > 0$. Conversely, they will be repelled if $2f_\kappa + 3f_\rho < 0$.

We proceed with the derivation of the  spin angular momentum for modes with $|n|>0$. 
Substituting \eqref{Psi} into \eqref{SingleSpin}, we find the only nonzero spin component is the axial one, which is given by
\begin{align}
    \label{Snm0}
        \tilde{S}_z^{nm} = 2n
        \left|A^{nm}R^{nm} \right|^2
        \frac{ [J_n^2\!\left(k_\varrho^{nm} \varrho\right)]'}{k_\varrho^{nm}\varrho}.
\end{align}
Modes with no vorticity (\(n=0\)) lack spin. On the other hand, eliminating the axial component of a normal mode induces spin in the \(z\) direction. A detailed examination of \eqref{Snm0}, using the expression \(\lim_{x \rightarrow 0} \left[{J_n^2(x)}\right]'/x = {\delta_{n,\pm 1}}/{2}\), indicates that only the \((\pm 1\,m)\) mode exhibits spin along the \(z\) axis.
Hence,
\begin{equation}
    \lim_{\varrho\rightarrow0}\tilde{S}_z^{\pm1m} = \pm \left|A^{1m}R^{1m} \right|^2.  
\end{equation}

The amplitudes of the the energy density and Gor'kov potential as well as the spin density
 can be obtained using \eqref{Delta_nml}, \eqref{GammaTotal}, and \eqref{Anml}.
It can be shown that
\begin{equation}
       |A^l|^2, |A^{nm}|^2 \sim \textit{Sh}^2= \left(R/\delta\right)^2.
\end{equation}
The parameter $\textit{Sh}=R/\delta$ is  known as the
shear wave number of the cavity~\citep{Tijdeman1975}.
This suggests that for decoupled axial and vortex resonant modes, the higher shear wave numbers lead to more efficient nonlinear acoustic phenomena,  counteracting the typical viscous dissipation.

\subsection{Acoustic dual modes}

One potential application of dual modes in cylindrical cavities is to levitate cells or microparticles, providing an additional degree of freedom: rotation around an axis of symmetry. The dual mode, composed of an axial mode \((0\,0\,l)\), ensures the formation of \(l\) levitation planes at the mid-height of the cavity, while circular modes \((n\,m\,0)\) are responsible for clustering particles and inducing a spin-torque on the trapped particles.
From \eqref{AR_dualmode}, we note that the cavity's aspect ratio to host the dual mode the $ (0\,0\,l) + ( n\, m\, 0) $  is
\begin{equation}
    \frac{R}{H} = \frac{j'_{nm}}{l \pi}.
\end{equation}
Moreover, according to \eqref{UDualMode} and \eqref{SDualMode},  the Gor'kov potential and  spin density of this mode
near the resonance $(\omega\approx\omega^{l}\approx \omega^{nm0})$
can be expressed by
\begin{subequations}
    \begin{align}
    \label{Unm_l}
            \tilde{U}_{l}^{nm} 
        &= \tilde{U}^{l} + 
                \tilde{U}^{nm}+ \tilde{U}_\text{cross},\\
        \tilde{\vec{S}}_{l}^{nm} 
        &=
            \tilde{{S}}^{nm}_z \vec{e}_z + \tilde{\vec{S}}_\text{cross}.
            \label{SpinTotal}
    \end{align}
\end{subequations}
The cross terms are calculated with \eqref{crossTermU} and \eqref{crossTermS}.
Accordingly, we have
\begin{subequations}
    \begin{align}
    \nonumber
        \tilde{U}_\text{cross}&= 
 A^lA^{nm}\left|R^{l}R^{nm}\right|
         J_{n_1}(\tilde{\varrho}^{nm})
         \biggl[(2f_\kappa - 3 f_\rho)
        \cos \!\left(n\varphi - \Delta \varphi_l^{nm}\right)\\
        &+
        3f_\rho \left(\frac{k_z^l}{k}\right)^2\cos \!\left(n\varphi + \Delta \varphi_l^{nm}\right)
        \biggr]\cos \tilde{z}^l,
        \label{Ucross_nm}
        \\
        \nonumber
        \tilde{\vec{S}}_{\text{cross}} &=
        4
         A^l A^{nm}\left|R^{l}R^{nm}\right|
        \biggl[
            n \cos (n \varphi+\Delta \varphi_l^{nm}) \frac{J_{n}(\tilde{\varrho}^{nm})}{\tilde{\varrho}^{nm}}
            \vec{e}_\varrho \\
            &-\sin (n \varphi+\Delta \varphi_l^{nm}) \, J_{n}'(\tilde{\varrho}^{nm})
            \,\vec{e}_\varphi
        \biggr]\sin^2 \tilde{z}^l,
        \label{Scross}
    \end{align}
\end{subequations}
where the additional phase is $\Delta \varphi_l^{nm}=\angle  (R^l)^*R^{n m}$.

The spin component in \ref{Scross} arises from the nonlinear interaction between two orthogonal modes, $(0\,0\,l)$ and $(n\,m\,0)$. This component can be transferred to a suspended particle, much smaller than the wavelength, resulting in an acoustic radiation torque on the particle~\citep{Zhang2014}. Consequently, the particle may rotate around one of its principal axes. 
Notably, even when $n=0$ a spin component is still present, which given by
\begin{align}
    \tilde{\vec{S}}_{\text{cross}} &=
    4 A^l A^{0m}\left|R^{l}R^{0m}\right| \sin (\Delta \varphi_l^{0m}) \, J_{1}(\tilde{\varrho}^{0m})
    \sin^2 \tilde{z}^l \,\vec{e}_\varphi.
\end{align}
This result demonstrates that, even in the absence of vortex excitation, the spin angular momentum can still be generated in acoustofluidic resonators. 
It is also worth noting that metallic nanorods have been observed to rotate in acoustofluidic cylindrical cavities with a nonvortex excitation at the base~\citep{Wang2012,Balk2014}. 
This rotational behavior can be attributed to the spin-torque generated by the nonlinear interaction of orthogonal modes within the cavity.

Finally, we note that the cross-terms also scale as 
$A^l A^{nm}\sim\textit{Sh}^2$. 
We thus conclude  that the amplitude of the Gor'kov potential and spin density of a dual  mode  also scales with shear wave number squared.

% The cross term $\tilde{U}_\text{cross}$ introduces introduces an azimuthal angular dependence in the Gor'kov potential through $\cos(n_1 \varphi)$.
% %
% To analyse the particle trapping points, which occur at the minimum of the potential $\tilde{U}_l^{nm}$, we need first to compute the Hessian matrix, which is defined as follows
% \begin{equation}
%     \mathcal{H}_l^{n_1m_1} = 
%     \begin{pmatrix}
%     \partial^2_{\varrho \varrho} & \partial^2_{\varrho\varphi} & \partial^2_{\varrho z} \\
%     \partial^2_{\varphi\varrho} & 
%     \partial^2_{\varphi\varphi}
%     & \partial^2_{\varphi z}\\
%     \partial^2_{z \varrho} & \partial^2_{z \varphi} U& \partial^2_{zz}
%     \end{pmatrix}
%     \tilde{U}_l^{n_1m_1}.
% \end{equation}
% To test if a point will trap a particle, we check the Hessian matrix is positive-definite at that point. 

%
% \begin{table}
% \centering
% \begin{tabular}{ccc}
% Aspect ratio & Dual vortex mode & Pressure  \\
% \vspace{-.2cm}
% \\
% $\num{0.58606} $ & $ (0\,0\,1) + ( 1\, 0\, 0) $ & $
% A^{001}  \cos\!\left(k^1_z z\right) +
%         A^{100}  J_1\!\left(k_\varrho^{10}\varrho\right) \ee^{\ii \varphi} $
%         \\
% $\num{0.97219} $ & $ (0\,0\,1) + ( 2\, 0\, 0) $ & $
% A^{001}  \cos\!\left(k^1_z z\right) +
%         A^{200}  J_2\!\left(k_\varrho^{20}\varrho\right) \ee^{2\ii \varphi} $
% \end{tabular}
% \caption{The dual vortex modes involving the first-axial mode. 
% The aspect ratio required to produce the dual mode configuration is  calculated from \eqref{dual-freqWavenumber}.
% \label{table:DualModes}}
% \end{table}

\section{Theoretical versus numerical results}

\begin{figure}
    \begin{center}
         \includegraphics[scale=.8]{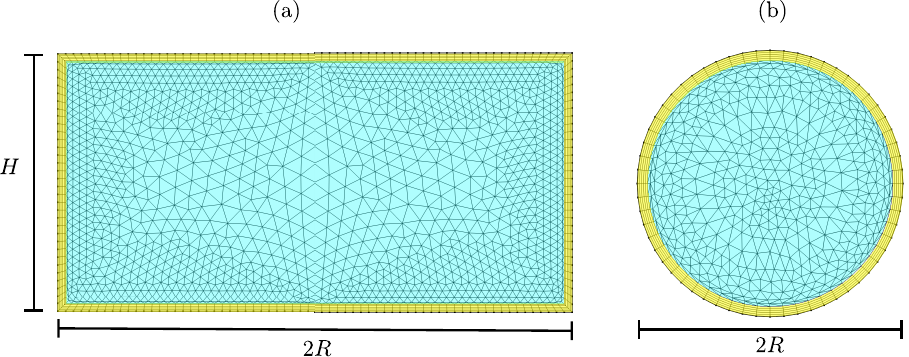} 
    \end{center}
    \caption{
    Visualization of the finite element simulation mesh.  (a) The lateral and (b) transverse cross-sectional view of the cavity. The mesh features the bulk region (blue) with a tetrahedral unstructured elements and the viscous boundary layer (yellow) near the walls. The illustrated meshes are not in scale.
    \label{fig:mesh}}
\end{figure}

The theoretical results are now compared with finite element simulations conducted in \textsc{Comsol} Multiphysics (Comsol AB, Sweden). Figure~\ref{fig:mesh} illustrates the geometry setup and mesh utilized in the simulations. 
Panel (a) depicts the lateral cross-sectional view of the model mesh, while panel (b) provides a cross-sectional view perpendicular to the cavity axis. 
The mesh in the fluid bulk (in blue color) is characterized by a tetrahedral unstructured pattern with less dense elements. 
Near the walls, the viscous boundary layer (in  yellow color) is delineated by a finer structured mesh designed to capture the short-range variation of fluid flow. 
To compute the pressure and velocity field within the fluid inside the resonator, the \emph{Thermoviscous Acoustics, Frequency Domain Interface} is employed. 
The simulations are executed with physical parameter given in table~\ref{tab:PhysicalProperties} and under the adiabatic hypothesis, as discussed in appendix~\ref{app:isentropic}.
The excitation velocity  at the cavity bottom has  a value of $|v_\text{ex}| =\SI{1}{\milli\meter\per\second}$. 
The radiation force is obtained for a polystyrene probe particle with acoustophoretic parameters $f_\kappa=0.3128$ and $f_\rho =0.0342$.
Finally, the simulations were conducted on a workstation featuring an Intel Xeon (R) E5-2690 processor with a $\SI{3}{\giga\hertz}$ clock, and equipped with $\SI{384}{\giga\byte}$ of memory.
To compare the theoretical to numerical results, we use the coefficient of determination $\mathcal{R}^2$, which measures
how well a theoretical model explains the variability in a numerical data. 
A value of $\mathcal{R}^2=1$ indicates that the theoretical model perfectly matches the data, whereas $\mathcal{R}^2=0$ suggests that the theoretical model does no better than predicting the mean of the numerical data.

\subsection{Single modes}
In the results for single modes, a water-filled microcavity with dimensions $H=R=\SI{250}{\micro\meter}$ and a volume of $V_0=\SI{49}{\nano\liter}$ is considered, representing a typical cavity in acoustofluidics devices.

\begin{figure}
    \begin{center}
    \includegraphics[scale=.75]{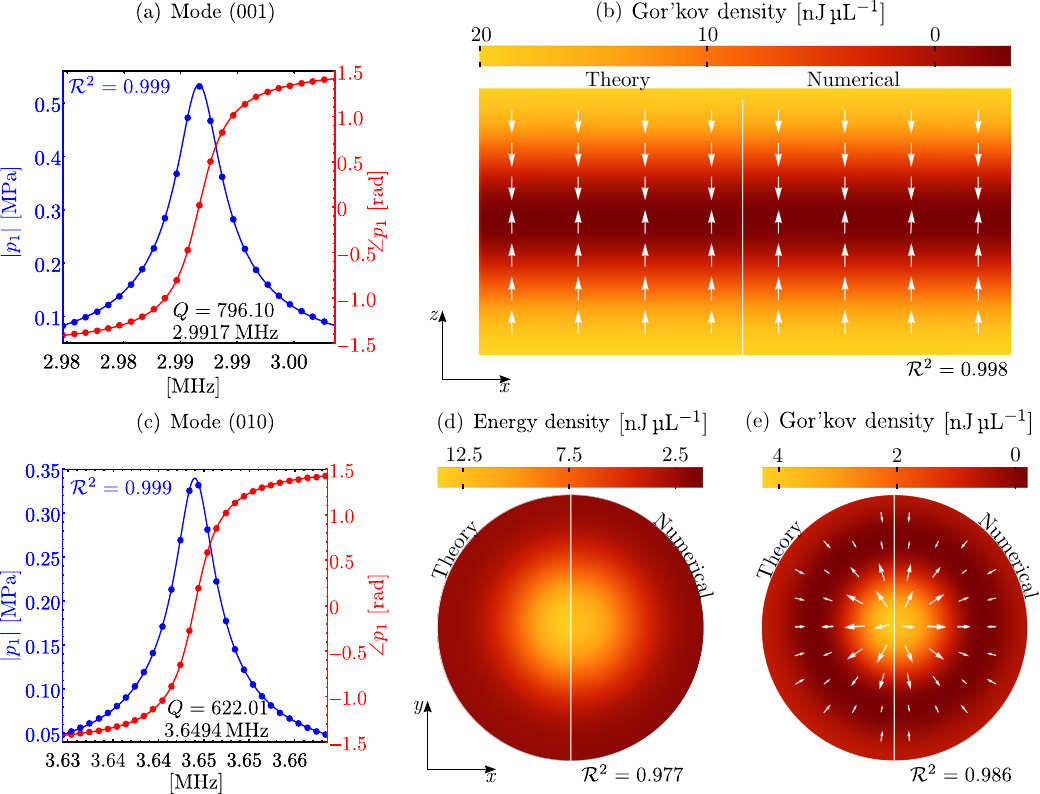} 
    \end{center}
    \caption{ 
    The $(0\,0\,1)$ and $(0\,1\,0)$ mode induced in a microcavity with $H=R=\SI{250}{\micro\meter}$.
    (a), (c)  The pressure-frequency plot (magnitude and phase) using \eqref{p1lm}, and the corresponding resonance frequency and $Q$ factor. 
    The blue and red dots indicate numerical results. 
    (b) The Gor'kov potential in the $xz$ plane (background image) obtained from \eqref{U001}, with the related radiation force field (white arrows) acting on a $\SI{10}{\micro\meter}$-polystyrene probe particle.
    (d) The energy density computed with \eqref{EnergyDensity_nm0} and (e) the Gor'kov potential for the probe particle calculate with \eqref{Unm0}.
    All fields are obtained at the corresponding resonance frequency.
    The vertical white lines delimit  the theoretical (left) and numerical (right) results.
    \label{fig:NonvortexModes} }
\end{figure}

Figure~\ref{fig:NonvortexModes} presents the $(0\,0\,1)$ and $(0\,1\,0)$ mode  induced in the microcavity.
The physical parameters of water used in this analysis are given in table~\ref{tab:PhysicalProperties}, while 
panels (a) and (c) plot pressure versus frequency as given in \eqref{p1lm}, with $Q=796.10$ and $622.01$, respectively. The resonance frequencies are $\SI{2.9917}{\mega\hertz}$ and $\SI{3.6494}{\mega\hertz}$.
Panel (b) depicts the Gor'kov potential density (background image) of the $(0\,0\,1)$ mode at the $xz$ plane, with white arrows illustrating the radiation force field. 
Probe particles will collect in the pressure node at the cavity's midheight.
Panels (d) and (e) show the energy density and Gor'kov potential of the $(0\,1\,0)$ mode. 
The related radiation force will trap particles in an annular region (in dark brown) in panel (e). 
The vertical white lines separate the theoretical (left) and numerical (right) results.
The numerical data and the theoretical predictions show excellent agreement  with $\mathcal{R}^2\sim 1$. 
It is important to mention that in the numerical simulations, the degrees of freedom (DOF) for the two-dimensional axisymmetric model used for the  $(0\,0\,1)$ and $(0\,1\,0)$ modes is $\num{288045}$.
Larger numbers of DOF typically require more computational resources, both in terms of memory and processing power.

In figure~\ref{fig:NonvortexModes}, panel (e), we note that the radiation force is zero in a ring with radius, denoted by $\varrho_\text{r}$, which is a solution of the transcendent equation derived by inserting \eqref{Unm0} into \eqref{RadiationForce}:
\begin{equation}
   3 f_\varrho J_2\!\left(k_\varrho^{01} \varrho_\text{r}\right)- (4 f_\kappa + 3 f_\varrho) J_0\!\left(k_\varrho^{01} \varrho_\text{r}\right) = 0.
\end{equation}
The numerical solution to this equation yields $\varrho_\text{r} = 0.6117 R$ or $\varrho_\text{r}=\SI{152}{\micro\meter}$, which is compatible with the particle trap region seen in panel (e).

\begin{figure}
    \begin{center}
    \includegraphics[scale=0.59]{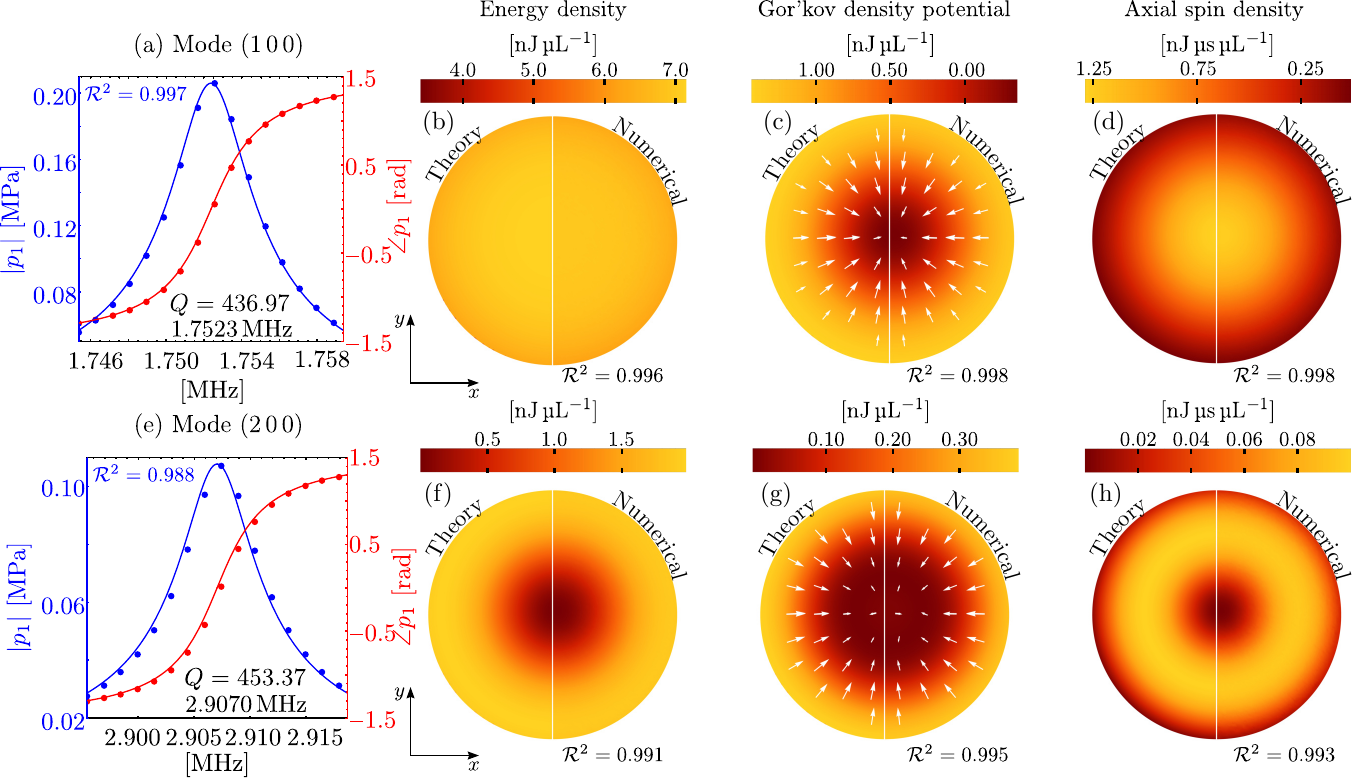} 
    \end{center}
    \caption{The $(1\,0\,0)$ and $(2\,0\,0)$  mode for the cavity
    described in figure~\ref{fig:NonvortexModes}.
    (a), (e) The acoustic pressure plotted against frequency using \eqref{p1lm}, along with the associated $Q$ factor and resonance frequency.
    The dots represent the numerical data.
    (b), (f) The energy density  calculated with \eqref{EnergyDensity_nm0}. (c), (g) Illustration of the Gor'kov potential for a $\SI{10}{\micro\meter}$-polystyrene probe particle using \eqref{Unm0}. 
    (d), (h) The spin angular momentum according to \eqref{Snm0}. 
    These fields are computed at corresponding resonance frequency of each mode.
    The vertical white lines separate the theoretical (left) and numerical (right) results.
    \label{fig:vortexModes}}
\end{figure}

In figure~\ref{fig:vortexModes}, we illustrate the $(1\,0\,0)$ and $(2\,0\,0)$  mode within the same cavity introduced in figure~\ref{fig:NonvortexModes}. 
Panels (a) and (c) present the pressure plotted against frequency using \eqref{p1lm}.
The associated $Q$ factor and resonance frequency are $436.97$ and $\SI{1.7523}{\mega\hertz}$, and $453.37$ and $\SI{2.9070}{\mega\hertz}$, respectively. 
The energy density given in \eqref{EnergyDensity_nm0} is shown in panels (b) and (f), while
the Gor'kov potential from \eqref{Unm0}  for a $\SI{10}{\micro\meter}$-polystyrene particle are seen in panels (c) and (g).
The  spin angular momentum as given by \eqref{Snm0} is shown in panels (d) and (h).
These modes exhibit no levitation plane, indicating their purely vortex nature. 
Notably, the inward radiation force field is responsible for concentrating particles in the central portion of the resonator without levitation. 
The spatial pattern of the axial spin differs significantly between these modes. 
In the $(1\,0\,0)$ mode, collected particles experience a positive spin along the $z$ axis, peaking at $\varrho=0$. 
Conversely, for the $(2\,0\,0)$ mode, the maximum axial spin, according to \eqref{Snm0}, occurs in a ring of radius $\varrho_\text{s}$, which is determined by the solution of
\begin{equation}
    \left[J_2^2\!\left(k_\varrho^{20} \varrho_\text{s}\right)\right]'' =0 .
\end{equation}
We find $\varrho_\text{s}=\num{0.6759}R$, yielding $\varrho_\text{s} = \SI{169}{\micro\meter}$.
The axial spin potentially may induces a counterclockwise rotation on particles mostly depending on their geometric asymmetry.
Note that the spin vanishes at $\varrho=0$.
In the numerical simulations, it was observed that the DOF of the three-dimensional model  for the $(1\,0\,0)$ mode was $\num{2705168}$. 
On the other hand, the DOF for the $(2\,0\,0)$ mode is $\num{2475000}$.
According to the coefficient of determination, $\mathcal{R}^2\sim 1$, and visual inspection, the
theoretical and numerical data show excellent agreement.

\subsection{Dual modes}

Here we consider the $(0\,1\,0)+(0\,0\,1)$, $(1\,0\,0)+(0\,0\,1)$ and $(2\,0\,0)+(0\,0\,1)$  mode,
as they correspond to simplest modes for trapping and inducing torque on suspended particles. 
To have develop these modes, the cavities aspect ratio should be, respectively, 
$R/H=\num{1.2197}, \num{0.5860}$, $\num{0.9721}$.
The radius of all cavities is fixed to $R=\SI{250}{\micro\meter}$.
Also, the cavities are filled with water.
\begin{figure}
    \begin{center}
    \includegraphics[scale=0.6]{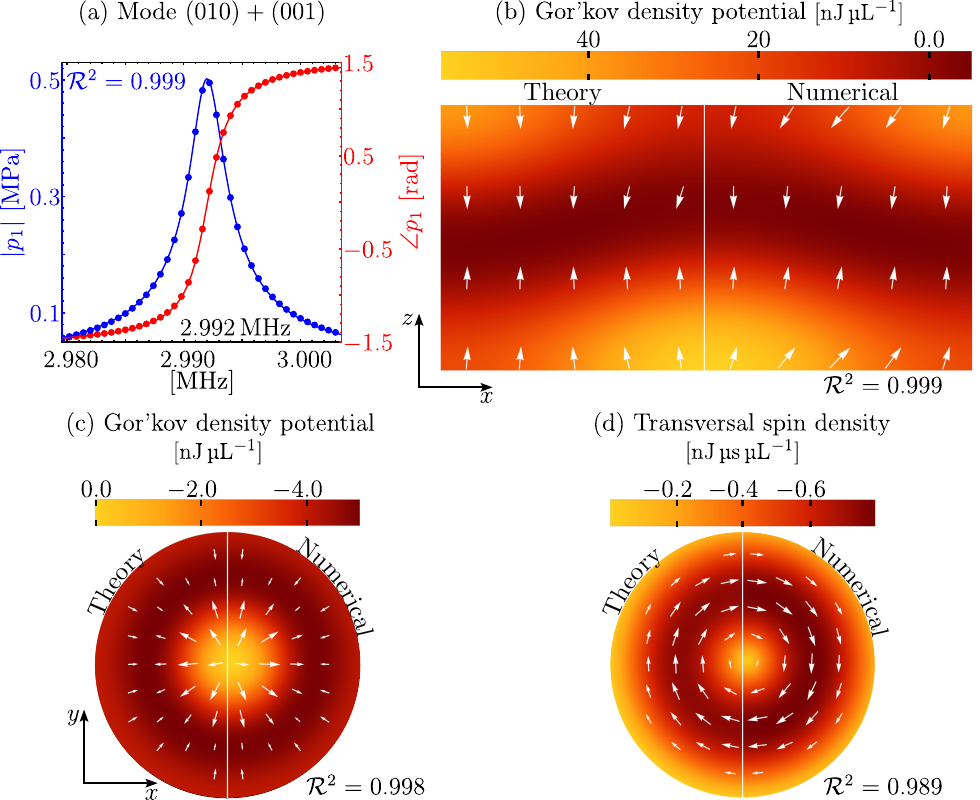} 
    \end{center}
    \caption{
    The  $(0\,1\,0)+(0\,0\,1)$ mode for a cavity with aspect ratio
   $R/H=\num{1.2197}$.
    Panel (a) shows the magnitude (in blue) and phase (in red) of the acoustic pressure  around the resonance frequency $\SI{2.992}{\mega\hertz}$, plotted using \eqref{p1lm}.
    The numerical data are represented by dots.
    Panel (b) presents the Gor'kov potential density for the probe particle, with white arrows depict the corresponding acoustic radiation force. 
    The theoretical part of the potential is calculated using \eqref{Unm_l}. 
    Panel (c) exhibits the Gor'kov potential at the cavity mid-height,
    while panels (d) illustrate the magnitude of the transverse  spin density, with white arrows denoting the related vector field. 
    The theoretical part of the spin is computed with \eqref{SpinTotal}. 
    The acoustic fields in panels (b), (c), and (d) are computed at the resonance frequency.
    The vertical white lines separate the theoretical (left) and numerical (right) results.
    \label{fig:DualModes010}}
\end{figure}

Figure~\ref{fig:DualModes010}, panel (a), displays the acoustic pressure for the $(0\,1\,0)+(0\,0\,1)$ mode near the resonance frequency of 
\(\SI{2.992}{\mega\hertz}\), based on numerical simulations with \num{331164} DOF. 
Panel (b) depicts the Gor'kov potential density in the $xz$ plane, along with the radiation force field shown by the white arrows. 
This potential is axisymmetric.  
Panel (c) shows the Gor'kov potential in the $xy$ plane at the mid-height of the cavity, where the radiation force field (white arrows) traps particles in an annular region around the cavity central axis. 
Panels (d) presents the transverse spin component, with white arrows representing the related vector field.
This spin arises due to the interaction between the transverse and axial velocity fields in the fluid. 
The spin  exhibits characteristics of a solenoidal field (circulation). 
The theoretical and numerical results are in excellent agreement, with \(\mathcal{R}^2 \sim 1\), as seen by visual inspection.

\begin{figure}
    \begin{center}
    \includegraphics[scale=0.6]{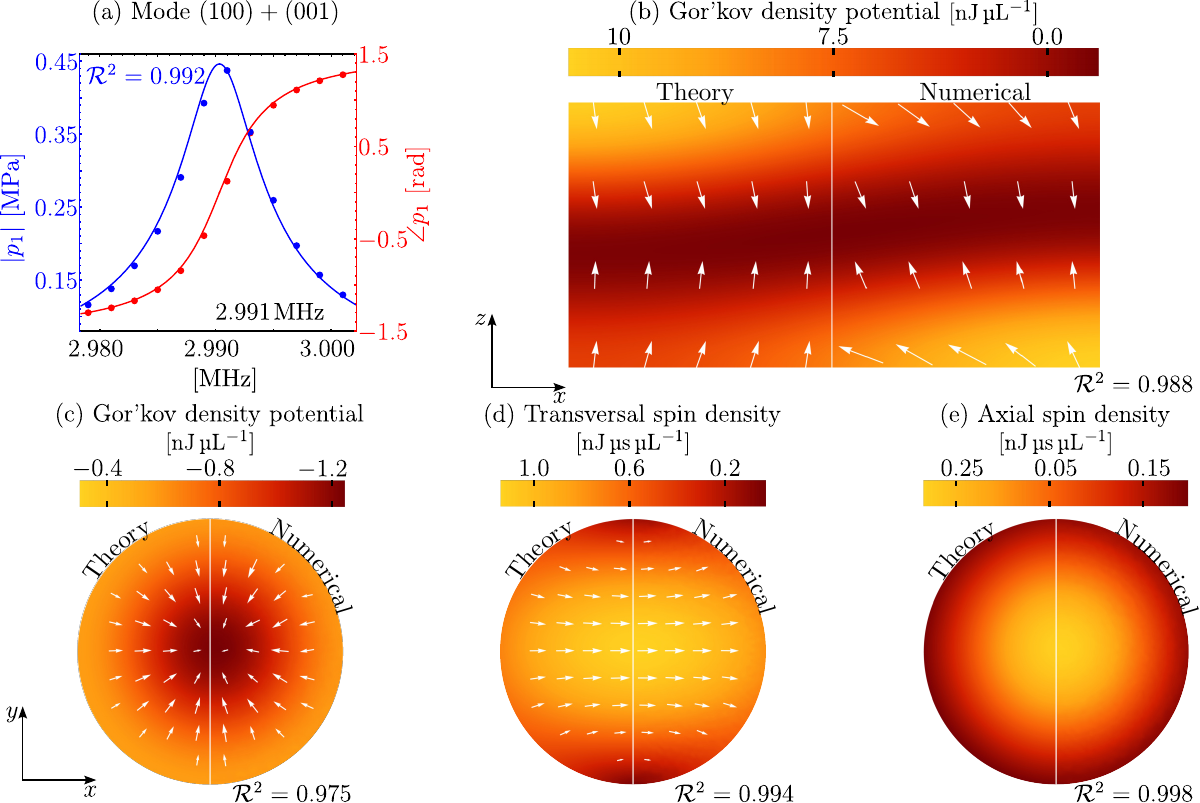} 
    \end{center}
    \caption{
    The  $(1\,0\,0)+(0\,0\,1)$ mode for a cavity with
   $R/H=\num{0.5860}$ and $R=\SI{250}{\micro\meter}$.
    Panel (a) show the acoustic pressure (magnitude in blue and phase in red)  around the resonance frequency $\SI{2.991}{\mega\hertz}$, and plotted against frequency using \eqref{p1lm}.
    The dots represent the numerical data.
    Panel (b) presents the background map is the Gor'kov potential density for the  probe particle. 
    The theoretical part is computed using \eqref{Unm_l}. 
    Panel (c) exhibit the Gor'kov potential at the mid-height,
    while panels (d) and (e) illustrate the magnitude of the transverse and axial components of the spin density at the mid-height, respectively. 
    The theoretical part of these components are calculated with \eqref{SpinTotal}. 
    The white arrows in panel (d) correspond to the spin vector field.
    The acoustic fields in are computed at the resonance frequency.
    The vertical white lines separate  theoretical (left) and numerical (right) results.
    \label{fig:DualModes100}}
\end{figure}

Figure~\ref{fig:DualModes100}, panel (a) shows the acoustic pressure for the  $(1\,0\,0)+(0\,0\,1)$ mode around the resonance frequency of 
\(\SI{2.991}{\mega\hertz}\).
A significant increase in the DOF 
(\num{1900422}) is noted, since this simulation requires a three-dimensional (3D) mesh.
Panel (b) presents the Gor'kov potential density in the $xz$ plane, along with the radiation force field (white arrows). 
The potential in this plane is asymmetric due to the cross term in \eqref{Ucross_nm}, which depends on the azimuthal angle $\varphi$. 
In contrast, panel (c) shows a symmetric Gor'kov potential (with $U_\text{cross}=0)$ in the mid-height plane, where the radiation force field directs particles towards the cavity's central axis. 
Panels (d) and (e) illustrate the transverse and axial spin components, respectively. 
The white arrows in panel (d) indicate the transverse spin field at mid-height, which results from the interaction between the transverse and axial fluid velocity fields. 
The axial spin component is most pronounced in the central region of the cavity but diminishes near the lateral walls, while the transverse spin is predominantly oriented left-to-right and is approximately four times stronger than the axial spin. A circulating pattern is observed in the transverse spin field, consistent with the behavior expected of a solenoidal field.
We see that theoretical and numerical results show excellent agreement, with \(\mathcal{R}^2 \sim 1\), confirmed by visual inspection.

\begin{figure}
    \begin{center}
    \includegraphics[scale=0.6]{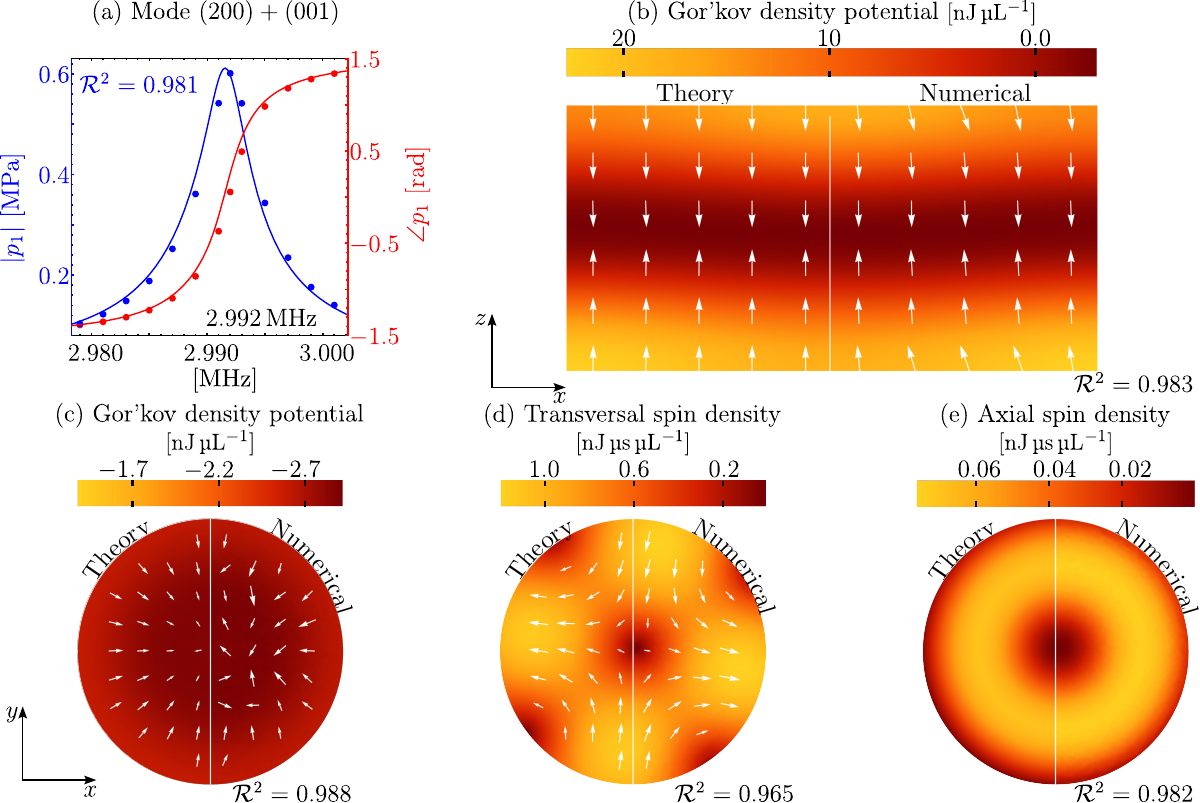} 
    \end{center}
    \caption{The  $(2\,0\,0)+(0\,0\,1)$ mode for a cavity with an aspect ratio of $R/H = \num{0.9721}$ and radius \( R = \SI{250}{\micro\meter} \). Panels (a) and (e) display the acoustic pressure around the resonance at \(\SI{2.992}{\mega\hertz}\), plotted as a function of frequency using \eqref{p1lm}, with numerical data shown as dots. Panel (b) depicts the Gor'kov potential density for a probe particle, calculated using \eqref{Unm_l}. Panel (c) shows the this potential at mid-height with the corresponding radiation force (white arrows), while panels (d) and (e) present the transverse and axial spin density magnitudes, calculated via \eqref{SpinTotal}. White arrows in panel (d) represent the spin vector field. The presented acoustic fields are computed at resonance, with vertical white lines dividing the theoretical (left) and numerical (right) results.
    \label{fig:DualModes200}
    }
\end{figure}

Figure~\ref{fig:DualModes200}, panel (a) displays the acoustic pressure for the $(2\,0\,0)+(0\,0\,1)$ mode near the resonance frequency of \(\SI{2.992}{\mega\hertz}\), with \num{1900422} DOF in the numerical simulations. 
Panel (b) shows the Gor'kov potential density and radiation force field (white arrows) in the $xz$ plane. 
Conversely, panel (c) illustrates the Gor'kov potential in the $xy$ plane at mid-height, with the radiation force (white arrows) pointing toward the cavity's axis of symmetry. 
The potential well is more shallow compared with the one in figure~\ref{fig:DualModes100}.
Moreover, the simulation results exhibit some numerical fluctuations in the computed radiation force field. These fluctuations can, in principle, be mitigated by increasing the mesh resolution, though this would require greater computational memory.
Panels (d) and (e) depict the transverse and axial spin components, respectively, with white arrows in panel (d) representing the transverse spin field at mid-height. 
The axial spin is more dominant in the annular region around the central axis.
The transverse spin show a circulating pattern around the for dark spots at the edge of the cavity.
Finally, theoretical and numerical results align closely, with \(\mathcal{R}^2 \sim 1\).

\section{Summary and Conclusions}

This study develops a detailed theoretical framework for understanding acoustic vortex modes in cylindrical resonator cavities, with particular attention to the role of viscous boundary layers. We employed the isentropic hypothesis to simplify the thermodynamic treatment by assuming negligible heat transfer within the fluid medium. This assumption is justified for acoustofluidic cavities with a low-amplitude excitation, where the induced temperature rise is expected to be much smaller than $\SI{1}{\kelvin}$--see Appendix~\ref{app:isentropic}. 

Our investigation explored both single- and dual-mode configurations in microcavity resonators. 
The single modes analyzed consisted of either pure axial or pure transverse modes, while dual modes involved a simultaneous combination of axial and transverse modes. 
Dual  modes are particularly noteworthy because they can exhibit spin angular momentum in both axial and transverse directions, offering new degrees of freedom for advanced manipulation of microparticles in acoustofluidic systems. 
Interestingly, even in the absence of vortex excitation, a transverse spin angular momentum can still be generated in acoustofluidic cylindrical resonators due to the nonlinear interaction between two orthogonal waves in the cavity. 
This result may provide an explanation for the observed rotation of metallic nanorods in acoustofluidic cavities~\citep{Wang2012, Balk2014}, with a nonvortex excitation at the base.

We observed that nonlinear acoustic fields near resonance scale with the square of the shear wave number, \( \textit{Sh}^2 = (R/\delta)^2 \). 
Additionally, the theoretical results show excellent agreement with finite element simulations, as indicated by the coefficient of determination \( \mathcal{R}^2 \sim 1 \) and visual inspection of the results. 
This consistency validates the accuracy of the theoretical approach and its potential for practical applications in describing acoustofluidic systems.
It is important to highlight that our hard-wall model for acoustofluidic cavity resonators establishes a theoretical benchmark for evaluating the performance limits of energy storage and particle manipulation efficiency in cavities constructed from practical materials such as glass and silicon. 

The numerical simulations of acoustic vortex modes in cylindrical resonators require a relatively high number of DOFs, typically in the low-million range. 
With this level of complexity, simulations demand significant computational resources, including increased memory capacity and processing power. 
Given the computational challenges, the development of accurate and efficient theoretical models is crucial for advancing the field of acoustofluidic resonators.

Experimental verification of the predicted spin angular momentum in cavity resonators is important to validate the theoretical results presented in this study. 
Experiments can focus on detecting the induced rotations in a probe particle due to the transfer of the spin angular momentum to the particle . 
While the theoretical model provides robust insights into the behavior of spin angular momentum in acoustofluidic systems, experimental confirmation is essential to bridge the gap between theory and applications.

Future research could expand upon this work by exploring more complex scenarios that include acoustic streaming and thermal interactions. 
Extending the model to account for such phenomena would provide a better understanding of the fluid mechanics of acoustic microcavities and enable the design of more advanced acoustofluidic systems for practical applications.

\appendix
\section{Isentropic hypothesis and heat generation}
\label{app:isentropic}
To investigate the hypothesis that no heat is diffused throughout the fluid medium, we consider the fluid dynamic equation for entropy per unit mass $(s)$,
\begin{equation}
    \rho T \frac{\dd s}{\dd t} = \left(\vec{\sigma} + p \mathsfbi{I} \right) : \grad {\vec{v}} + \nabla \cdot (\kappa \nabla T),
    \label{heatEquation}
\end{equation}
where $\dd/\dd t$ is the material derivative, the symbol $:$ denotes the double-dot product. 
In the first-order approximation, this equation simplifies to
\begin{equation}
    \rho_0 T_0 \partial_t s_1 = k_\text{th,0} \nabla^2 T_1,
    \label{linearEntropy}
\end{equation}
where $k_\text{th,0}$ represents the thermal conductivity. 
The subindex $0$ indicates the parameter does not depend on other thermodynamic variables, i.e., it is constant.
Expressing entropy as a function of pressure and temperature, $s = s(p,T)$, we have the entropy differential
\begin{equation}
    \dd s = \left(\frac{\partial s}{\partial p}\right)_{T} \dd p + \left(\frac{\partial s}{\partial T}\right)_{p} \dd T.
    \label{ds}
\end{equation}
This leads to the introduction of the isothermal compressibility $\kappa_{T}$, the isobaric thermal expansion coefficient $\alpha_{p}$, and the specific heat capacity at constant pressure $c_{p}$,
\begin{equation}
    \kappa_{T} = \frac{1}{\rho} \left(\frac{\partial \rho}{\partial p}\right)_{T}, \quad
    \alpha_{p} = -\frac{1}{\rho} \left(\frac{\partial \rho}{\partial T}\right)_{p}, \quad
    c_{p} = T \left(\frac{\partial s}{\partial T}\right)_{p}.
\end{equation}
The quantity $\left(\frac{\partial s}{\partial p}\right)_{T}$ follows from the differential of the Gibbs potential,
\begin{equation}
    \dd \!\left (u - Ts + \frac{p}{\rho}\right) = \frac{\dd p}{\rho} - s \,\dd T,
\end{equation}
which implies the Maxwell relation
\begin{equation}
    \left(\frac{\partial s}{\partial p}\right)_{T} = - \left(\frac{\partial }{\partial T}\frac{1}{\rho}\right)_{p} = \frac{1}{\rho^{2}} \left(\frac{\partial \rho}{\partial T}\right)_{p} = -\frac{\alpha_p}{\rho}.
\end{equation}
In the first-order approximation, the entropy differential in \eqref{ds} becomes
\begin{equation}
    \dd s_1 = -\frac{\alpha_{p,0}}{\rho_0} \, \dd {p}_1 + \frac{c_{p,0}}{T_0}\, \dd T_1.
    \label{ds1}
\end{equation}
In the linear approximation, $\dd/\dd t=\partial_t$, which
 immediately implies
\begin{equation}
    \partial_t s_1 = -\frac{\alpha_{p,0}}{\rho_0}\,  \partial_t {p}_1 + \frac{c_{p,0}}{T_0}\,\partial_t T_1.
    \label{partialS1}
\end{equation}
Substituting this result into the expression for \eqref{linearEntropy}, we obtain
\begin{equation}
    \partial_t T_1 = D_{\text{th},0} \, \nabla^2 T_1 +  \frac{\alpha_{p,0} T_0 }{\rho_0 c_{p,0}} \, \partial_t {p}_1,
    \label{PressureTemperatureDiffusivity}
\end{equation}
where $D_{\text{th},0} = k_\text{th,0}/\rho_0 c_{p,0}$ is the thermal diffusivity.
For water at $\SI{25}{\degreeCelsius}$, the thermal diffusivity is $D_{\text{th},0} = \SI{1.4548e-7}{\meter\squared\per\second}$.

In the presence of a harmonic wave, the timescale for temperature and pressure variation is governed by the wave period $\omega^{-1}$.
Therefore, the Laplacian of the temperature in the cavity bulk follows $\nabla^2 T_1 \sim k^{2} T_1$. 
Incorporating this approximation into \eqref{PressureTemperatureDiffusivity}, we derive the isentropic condition as $ k D_{\text{th},0}/c_0 \ll 1$.
This condition dictates that the acoustic wavelength should satisfy
\begin{equation}
    \lambda \gg \frac{2 \upi D_{\text{th},0}}{c_0}.
\end{equation}
For water at $\SI{25}{\degreeCelsius}$, the acoustic wavelength must satisfy $\lambda\gg \SI{0.1}{\nano\meter}$, a condition typically met in acoustofluidic settings. 
Consequently, we can infer that the first-order temperature is primarily dependent on pressure in the fluid bulk:
\begin{equation}
\partial_t T_1 \approx \frac{\alpha_{p,0} T_0}{\rho_0 c_{p,0}}, \partial_t {p}_1.
\end{equation}
This implies that the acoustic phenomena in the bulk follows an isentropic process, where $T_1 = T_1(\tilde{p}_1, s_0)$ and $s_0$ represents a constant reference entropy. Additionally, the pressure-density relation in \eqref{state_equation} can be reliably assumed within the acoustofluidic cavity under examination.

Referring to \eqref{heatEquation}, it is noted that the predominant mechanism for heat generation (or power loss density) within the viscous boundary layer arises from  the mechanical work of viscous stress forces, represented by $\eta_0 |\grad {\vec{v}}\!:\!\grad {\vec{v}}|$.
This phenomenon, a second-order effect, contributes to the thermal changes of the system.
Referring to \eqref{v1cnml}, the time average of the total power dissipated corresponds to the mean  power loss density times the cavity volume,  which scales as $P_\text{loss}= \pi R^2 H\eta_0\Mach^2 c_0^2/2(\delta\Gamma k H)^2$, at resonance. 
In a stationary regime, this dissipated power is counterbalanced by the power conducted through the cavity walls.
When considering dissipation across the boundaries, the dissipated power, $P_\text{diss}$, can be approximated by $P_\text{diss}=- 2\pi R^2 (1+H/R) k_\text{th,0} \nabla T \approx$, but $ \nabla T \approx  \Delta T /\delta$ (across the boundaries).
Hence, from the relation $P_\text{diss} = P_\text{loss}$, we find
the resultant temperature change $\Delta  T \approx \Mach^2 H\eta_0 c_0^2/[4\delta k_\text{th,0} (1+H/R)]$.
For water at $\SI{25}{\degreeCelsius}$, we have
$
\Delta T \approx (\SI{821.81}{\kelvin})\Mach^2 (\Gamma k H)^{-2} (1+H/R)^{-1} (H/\delta)$.
In typical acoustofluidic settings, $H/\delta \sim 100$, $H/R\sim 1$, $k H\sim \pi$, and $\Gamma\sim100$. 
Hence, a temperature changes of about  $\SI{1}{\kelvin}$ requires   a high-amplitude excitation of the order of $\Mach\sim10^{-2}$.
The excitation used in acoustofluidic cavities are much smaller than this one. 
So one expects that the temperature rise in the cavity will be much smaller than $\SI{1}{\kelvin}$.
Such variation can induce up to a $3\%$ alteration in the shear and bulk viscosity at $\SI{25}{\degreeCelsius}$ \citep{Hahn2015a}, with negligible effects in other physical parameters. 

An additional consideration is the attenuation of acoustic waves in the vicinity of walls due to the presence of a thermal boundary layer, which arises from the oscillating temperature field. 
As delineated by \cite{Hahn2015a}, the thermal damping in cavities with  silicon walls--a representative hard material--is approximately $250$ times less significant than its viscous counterpart. 
Finally, the modest impact caused by a small temperature rise suggests that the simplification of the model, omitting detailed thermal conduction analysis, remains viable within the context of the current analysis.

\section{Derivation of boundary conditions}
\label{app:BC}

We assume the linear shear velocity $\tilde{\vec{v}}_\text{s}$ is highly attenuated with distance to a wall.
This velocity is then decomposed as
\begin{equation}
    \label{shear_decomp}
    \tilde{\vec{v}}_\text{s} =
    \begin{cases}
        A(z)\tilde{\vec{v}}^\text{b}_\text{s}(\varrho,\varphi), \quad & 0\le z \lesssim \delta,\\
        B(z)\tilde{\vec{v}}^\text{t}_\text{s}(\varrho,\varphi), \quad & H-\delta \lesssim z \le H,\\
        C(\varrho)\tilde{\vec{v}}^\text{w}_\text{s}(\varphi,z), \quad &  R-\delta \lesssim \varrho \le R.
    \end{cases}
\end{equation}
This vector has components $\tilde{\vec{v}}_\text{s}=(v_{\text{s},\varrho},v_{\text{s},\varphi},v_{\text{s},z})$.
The functions $A$, $B$, and $C$ are the dimensionless velocity amplitudes.
Whereas $\tilde{\vec{v}}^\text{b}_\text{s}$,
$\tilde{\vec{v}}^\text{t}_\text{s}$,
and $\tilde{\vec{v}}^\text{w}_\text{s}$ describe  the shear velocity  field, respectively, near the bottom, top, and lateral wall of the microcavity.
We assume that the amplitude variation along the perpendicular direction is much stronger than in the transverse counterpart.
As a consequence, we have
\begin{equation}
    \nabla^2
    \tilde{\vec{v}}_\text{s} \approx
    \begin{cases}
        \left(\partial_z^2 A_z\right)\tilde{\vec{v}}^\text{b}_\text{s}(\varrho,\varphi), \quad &0\le z \lesssim \delta,\\
        \left(\partial_z^2 B_z\right)\tilde{\vec{v}}^\text{t}_\text{s}(\varrho,\varphi), \quad &H-\delta \lesssim z \le H,\\
        \left(\partial_\varrho^2 A_\varrho\right)\tilde{\vec{v}}^\text{w}_\text{s}(\varphi,z), \quad & R-\delta \lesssim \varrho \le R.
    \end{cases}
    \label{nabla_vs}
\end{equation}
The boundary condition
on the perpendicular amplitude functions is
\begin{equation}
    A_z(0)= B_z(H)= A_\varrho(R)= 1.
\end{equation}
We also require the shear amplitude functions decrease away from the walls.
Hence, substituting  \eqref{nabla_vs} into  \eqref{Helmholtz_vs} 
gives
\begin{equation}
    \label{v1s}
    A_z = \ee^{\ii k_\text{s} z}, \quad 
    B_z = \ee^{\ii k_\text{s} (H-z)}, \quad 
    A_\varrho = \ee^{\ii k_\text{s}(R-\varrho)}.
\end{equation}
Since the shear flow is incompressible,
$\diverg{\tilde{\vec{v}}_\text{s}} = 0$,
we see from \eqref{shear_decomp} and  \eqref{v1s} that
\begin{subequations}
    \label{v1sz0}
    \begin{align}
        v_{\text{s},z}|_{z=0} &=
         \frac{\ii}{k_s} \grad\cdot\tilde{\vec{v}}^\text{b}_\text{s},\\
         v_{\text{s},z}|_{z=H} &=
         -\frac{\ii}{k_s} \grad\cdot\tilde{\vec{v}}^\text{t}_\text{s},\\
        v_{\text{s},\varrho}|_{\varrho=R} &= -\frac{R}{1- \ii k_s R} \grad\cdot\tilde{\vec{v}}^\text{w}_\text{s}.
    \end{align}
\end{subequations}
This means the incompressibility ties the axial
shear velocity to what diverges from the transverse shear velocity fields.
Referring to the no-slip boundary condition in  \eqref{noslip} and noting that $\tilde{\vec{v}}_\text{c}=(v_{\text{c},\varrho},v_{\text{c},\varphi}, v_{\text{c},z})$, we find 
\begin{subequations}
    \label{noslipZ}
    \begin{align}
         {v}_{\text{c},z}|_{z=0}
        &= {v}_\text{b}-v_{\text{s},z}|_{z=0} = {v}_\text{b}-
         \frac{\ii}{k_s} \grad\cdot\tilde{\vec{v}}^\text{b}_\text{s},\\
        {v}_{\text{c},z}|_{z=H}
        &= -v_{\text{s},z}|_{z=H}=\frac{\ii}{k_s} \grad\cdot\tilde{\vec{v}}^\text{t}_\text{s},\\
        \label{v1c_rho}
        {v}_{\text{c},\varrho}|_{\varrho=R}
        &=-v_{\text{s},\varrho}|_{\varrho=R}= \frac{R}{1- \ii k_s R} \grad\cdot\tilde{\vec{v}}^\text{w}_\text{s}
        \approx
        \frac{\ii}{k_s} \grad\cdot\tilde{\vec{v}}^\text{w}_\text{s}.
    \end{align}
\end{subequations}
The last equation here requires $\delta/R\ll 1$. 
Because there is no shear movement on the cavity's boundaries, we have
\begin{subequations}
\label{shear_velocities}
    \begin{align}
        \tilde{\vec{v}}^\text{b}_\text{s} +  \left.\left(\tilde{\vec{v}}_\text{c}-{v}_{\text{c},z}
        \vec{e}_z\right)
        \right|_{z=0}
        &= \vec{0},\\
       \left. \tilde{\vec{v}}^\text{t}_\text{s}
       + \left(\tilde{\vec{v}}_\text{c} - {v}_{\text{c},z}
    \vec{e}_z\right)\right|_{z=H}
        &= \vec{0}
        ,\\
        \left.\tilde{\vec{v}}^\text{w}_\text{s}
        +\left(\tilde{\vec{v}}_\text{c} - {v}_{\text{c},\varrho}
        \vec{e}_\varrho\right)\right|_{\varrho=R}
        &= \vec{0}.
    \end{align}
\end{subequations}
Taking the  divergence $\diverg{}$ of these equations
brings us to
\begin{subequations}
    \begin{align}
        \diverg{\tilde{\vec{v}}}^\text{b}_\text{s}
        &=- \diverg{\left[\left.\left(\tilde{\vec{v}}_\text{c} - {v}_{\text{c},z}
        \vec{e}_z\right)\right|_{z=0} \right]}=
        -\left.\left(\diverg{\tilde{\vec{v}}}_\text{c}
        -\partial_z{v}_{\text{c},z}\right)\right|_{z=0} 
        ,\\
        \diverg{\tilde{\vec{v}}}^\text{t}_\text{s}
        &=- \diverg{\left[\left.\left(\tilde{\vec{v}}_\text{c} - {v}_{\text{c},z}
        \vec{e}_z\right)\right|_{z=H} \right]}=
        -\left.\left(\diverg{\tilde{\vec{v}}}_\text{c}
        -\partial_z{v}_{\text{c},z}\right)\right|_{z=H},\\
         \diverg{\tilde{\vec{v}}}^\text{w}_\text{s}
        &= -\diverg{\left[\left.\left(\tilde{\vec{v}}_\text{c} - {v}_{\text{c},\varrho}
        \vec{e}_\varrho\right)\right|_{\varrho=R}\right]}
        =
        -\left.\left[\diverg{\tilde{\vec{v}}}_\text{c}
        -\frac{1}{\varrho}\partial_\varrho(\varrho {v}_{\text{c},\varrho})\right]\right|_{\varrho=R}.
        \end{align}
\end{subequations}
Replacing these equations into \eqref{noslipZ} leads to
\begin{subequations}
    \begin{align}
         {v}_{\text{c},z}|_{z=0}
        & = {v}_\text{b} +
         \frac{\ii}{k_s} \left.\left(\diverg{\tilde{\vec{v}}}_\text{c}
        -\partial_z{v}_{\text{c},z}\right)\right|_{z=0},\\
        {v}_{\text{c},z}|_{z=H}
        &= -\frac{\ii}{k_s} \left.\left(\diverg{\tilde{\vec{v}}}_\text{c}
        -\partial_z{v}_{\text{c},z}\right)\right|_{z=H},\\
        \label{v1c_rho}
        {v}_{\text{c},\varrho}|_{\varrho=R}
        &=-
        \frac{\ii}{k_s}\left.\left[\diverg{\tilde{\vec{v}}}_\text{c}
        -\frac{1}{\varrho}\partial_\varrho(\varrho {v}_{\text{c},\varrho})\right]\right|_{\varrho=R}.
    \end{align}
\end{subequations}
Finally, we find the boundary conditions for the linear pressure using $\diverg{\tilde{\vec{v}}_\text{c}} = \ii k \tilde{p}_1$ and $\tilde{\vec{v}}_\text{c}\approx -\ii k^{-1}\tilde{p}_1$, 
\begin{subequations}
    \begin{align}
    \label{dzp0}
    \partial_z \tilde{p}_1
    &= 
     \ii k 
    v_\text{b}
    -
    \frac{\ii}{k_\text{s}} \left(\partial_z^2 + k^2\right) \tilde{p}_1,
    \quad z=0, \\
        \label{dzpH}
        \partial_z \tilde{p}_1
            &=
        \frac{\ii}{k_\text{s}} \left(\partial_z^2 + k^2\right) \tilde{p}_1,
        \quad z=H,\\
        \label{drpR}
        \partial_\varrho \tilde{p}_1
        &=
        \frac{\ii}{k_\text{s}} \left[\frac{1}{\varrho}\partial_\varrho(\varrho \partial_\varrho) + k^2\right] \tilde{p}_1, \quad \varrho=R.
    \end{align}
\end{subequations}
Here terms of order $\bigO\!\left(\Gamma_\text{c}\right)=\bigO(\epsilon^2)$ are neglected.
By noting that $\ii k_\text{s}^{-1}=(1+\ii) \epsilon k^{-1}$, we can write 
the boundary conditions as
\begin{subequations}
    \begin{align}
         \left[\partial_z
        + \epsilon \frac{1+\ii}{2k}\left(\partial_{z}^2 + k^2\right)\right] \tilde{p}_1\biggr|_{z=0} &= 0,
        \\
         \left[\partial_z
        - \epsilon \frac{1+\ii}{2k}\left(\partial_{z}^2 + k^2\right)\right] \tilde{p}_1\biggr|_{z=H} &= 0,
        \\
        \left\{ \partial_\varrho
        - \epsilon \frac{1+\ii}{2k}
       \left[\frac{1}{\varrho}\partial_\varrho (\varrho\partial_\varrho) + k^2\right]
\right\}\tilde{p}_1\biggr|_{\varrho=R} &= 0.
    \end{align}
\end{subequations}

\section{Circular wavenumber}
\label{app:radial}
By applying the boundary condition operator defined in  \eqref{R-BC} on the pressure in  \eqref{p1_sol}, we obtain
the equation satisfied by the circular wavenumber $k_\varrho$,
\begin{equation}
    k_\varrho R\, J_n'(k_\varrho R) -
    \frac{(1+\ii) \epsilon}{2k R}
    \biggl[(k_\varrho R)^2 J_n''(k_\varrho R)
     +
    k_\varrho R\,
    J'_n(k_\varrho R)+(k R)^2 J_n(k_\varrho R)
    \biggr]=0.
    \label{eqkr}
\end{equation}
The prime symbol means differentiation with respect to the function's argument.
Using the Bessel differential equation
$x^2J''_n(x)+xJ'_n(x)+(x^2-n^2)J_n(x)=0$, we reduce  \eqref{eqkr} to
\begin{equation}
    \label{mixed_BC}
     J_n'(k_\varrho R)= \left(1+\ii\right)
      \frac{\epsilon}{2k R}
 \frac{n^2 + (k R)^2-(k_\varrho R)^2}{ k_\varrho R}  J_n(k_\varrho R).
\end{equation}
Thus, we
may keep our analysis to a weak deviation from the inviscid Neumann boundary condition, $J_n'(k_\varrho R) =0$.
So for a given $(n\,m\,l)$ resonance mode, we can expand the circular wavenumber as
the sum of the inviscid part plus a small deviation term $\Delta {k}_{\varrho}^{nml}$,
\begin{equation}
\label{kr_decomp}
    k_\varrho^{nml} =
        k_{\varrho}^{nm} +
        \Delta {k}_{\varrho}^{nml},
        \quad \text{with}\quad 
        \left|
        \Delta {k}_{\varrho}^{nml}R
        \right| \ll1.
\end{equation}
After inserting $k_\varrho=k_\varrho^{nml}$ and $k = k_0^{nml}$ into  \eqref{mixed_BC}, expanding the result around
$\Delta {k}_{\varrho}^{nml} R = 0$, and keeping terms of order $\epsilon$,
we obtain 
\begin{align}
    \label{BCr_expanded}
     \Delta {k}_{\varrho}^{nml}  
 =(1+\ii) \frac{\epsilon^{nml}}{2 k R} \left[n^2 + \left(k_{z}^l R\right)^2
    \right] \frac{J_n(k_{\varrho}^{nm}R)}{(k_{\varrho}^{nm} R)^2\,J_n''(k_{\varrho}^{nm} R)} k_{\varrho}^{nm}.
\end{align}
We have used the fact that $J_n'(k_\varrho^{nm}R) = 0$.
Inserting 
$(k_\varrho^{nm}R)^2J''_n(k_\varrho^{nm}R)=[n^2 -(k_\varrho^{nm}R)^2]J_n(k_\varrho^{nm}R)$, $k^2-(k_\varrho^{nm})^2\approx (k_z^l)^2$
into  \eqref{BCr_expanded}, we arrive at
\begin{equation}
    \label{Deltanm}
    \Delta {k}_{\varrho}^{nml} = 
    (1+\ii)\frac{\epsilon^{nml}}{2k_0^{nml} R}
    \frac{n^2 + (k_{z}^l R)^2}{n^2-(k_{\varrho}^{nm} R)^2}  k_{\varrho}^{nm}.
\end{equation}

The $(0\,0\,l)$ mode corresponds to the absence of the radial and vortex modes, e.g., $k_\varrho^{00}=0$.
Only the axial mode is present.
Importantly, the no-slip boundary condition on the lateral wall $\varrho = R$, as given in  \eqref{v1c_rho}, implies that 
the axial shear mode $\tilde{\vec{v}}^\text{s}_\text{w}(\varphi,z)$ is converted into the compressional radial mode $v_{1,\varrho}^\text{c}$, which propagates in the microcavity  with a small amplitude.
So, we assume that 
\begin{equation}
    k_\varrho = \Delta k_\varrho^{00l}.
\end{equation}
Inserting this wavenumber into  \eqref{mixed_BC}, expanding it
around $\Delta k_\varrho^{00l}=0$ to the second-order along with $J_0(0)=1$, $J_0'(0)=0$, $J_0''(0)=-1/2$, 
$(\Delta k_\varrho^{00l})^2\epsilon\approx 0$,
$k_\text{c}^2 \epsilon \approx  (k_0^{00l})^2\epsilon$;
we obtain to the order $\epsilon$,
\begin{equation}
    \Delta k_\varrho^{00l}=\pm   \ii \sqrt{\left(1 + \ii\right)\frac{\epsilon^{00l}}{k_0^{00l} R}} k_0^{00l},
    \label{Deltanm2}
\end{equation}

In the end, we can write the circular wave number combining   \eqref{Deltanm} and \eqref{Deltanm2},
\begin{equation}
   \Delta {k}_{\varrho}^{nml} =
   \begin{cases}
   \pm  \ii \sqrt{\left(1 + \ii\right)\frac{\epsilon^{00l}}{k_0^{00l} R}} k_0^{00l},\quad &n,m=0,\\
   (1+\ii)\frac{\epsilon^{nml}
     }{2k_0^{nml}R}
    \frac{n^2 + (k_{z}^l R)^2 }{n^2-(k_{\varrho}^{nm} R)^2}  k_{\varrho}^{nm}, \quad &\text{otherwise}.
  \end{cases}
  \label{app:deltakr}
\end{equation}
%
% When the inviscid axial mode is suppressed ($l=0$),
% this equation reduces to
% \begin{align}
%     \nonumber
%   \Delta {k}_{\varrho}^{nm0} &=- (1+\ii)\epsilon^{nm0} \frac{
%      \,H}{2 R}
%      \left[ 
%     \frac{\left(q_0^{nm0} R\right)^2 }{ (j_{nm}')^2-n^2} - 1\right] q_\varrho^{nm}\\
%     &=  
%     - (1+\ii)
% \epsilon^{nm0} \frac{H}{2R}
%      \left[ 
%     \frac{n^2 }{ (j_{nm}')^2-n^2} \right]
%     \,  q_\varrho^{nm}.
%   \label{app:deltakr2}
% \end{align}
% Note that $q_0^{nm0} R= q_\varrho^{nm} R = j_{nm}'$.
% %
% %This term vanishes for nonvortex modes, $   \Delta {k}_{\varrho}^{0m0} = 0$.

\section{Pressure coefficient}
\label{app:PressureCoefficients}

Let us consider the case $l=0$ which leads to $k_{z}^0=0$.
So expanding \eqref{a1General} around $ k_z H = \Delta k_z^{nm0} H=0$ to the lowest order, we find
\begin{equation}
    \label{aznm0}
        a_1^{nm0} = 
        \frac{\ii k_0^{nm0}}
        {
        (\Delta k_z^{nm0})^2 H + (1+\ii)\, \epsilon^{nm0} k_0^{nm0}
        }.
\end{equation}
We can replace  \eqref{deltakz0} into  \eqref{aznm0} and use $c_0 = \omega^{0ml}/k_0^{0ml}$  to obtain 
\begin{equation}
    a_1^{nm0} = \frac{\ii c_0}{2H}
    \frac{1}{
         \omega-\omega^{nm0} -  \left[ \Gamma_\delta^{nm0}  - \frac{\epsilon^{nm0}}{k_0^{nm0} H}
         +
     \frac{ 2\Delta k_\varrho^{nm0}}{k_0^{nm0}} -\ii \left( \frac{\epsilon^{nm0}}{k_0^{nm0} H} +
     \Gamma_\text{c}^{nm0}\right)\right] \frac{\omega^{nm0}_0}{2}
        }.
        \label{aznm02}
\end{equation}
At the resonance $\omega=\omega^{nm0}$, 
the pressure amplitude reaches its maximum value, which implies that
the real part of the denominator of $a_1^{nm0}$ should vanish.
Hence, one obtains the relation
\begin{equation}
     \Gamma_\delta^{nm0}  +\frac{2\re[ \Delta k_\varrho^{nm0}]}{k_{0}^{nm0}} -  \frac{\epsilon^{nm0}}{k_{0}^{nm0} H}   = 0.
     \label{Deltak3}
\end{equation}
By noticing that $k_{0}^{nm0}=k_\varrho^{nm}$, we can re-write this for later convenience as
\begin{equation}
     \Gamma_\delta^{nm0}  +\frac{2k_\varrho^{nm}\re[ \Delta k_\varrho^{nm0}]}{\left(k_{0}^{nm0}\right)^2} -  \frac{\epsilon^{nm0} (k_\varrho^{nm})^2} {(k_0^{nm0})^3H}  = 0.
     \label{Deltak2}
\end{equation}

From  \eqref{Deltakrnml}, we note that
$\re[ \Delta k_\varrho^{nm0}] = \im[ \Delta k_\varrho^{nm0}]$,
which  allows us to write  \eqref{aznm02} as
\begin{align}
\nonumber
 &a_1^{nm0} =\frac{\ii c_0}{2H}\\
    &\times \frac{1}{
         \omega-\omega^{nm0} -  \left[ \Gamma_\delta^{nm0}  - \frac{\epsilon^{nm0}}{k_0^{nm0} H}
         +
     \frac{ 2\re[\Delta k_\varrho^{nm0}]}{k_\varrho^{nm}} -\ii \left( \frac{\epsilon^{nm0}}{k_0^{nm0} H} 
      -\frac{ 2\re[\Delta k_\varrho^{nm0}]}{k_\varrho^{nm}}
     + \Gamma_\text{c}^{nm0}\right)\right] \frac{\omega^{nm0}_0}{2}}
%         }.\\
% \nonumber
%     a_1^{nm0} &= \frac{c_0}{2H}
%      \left[
%          \left(\omega-\omega^{nm0} \right) -  \left( \Gamma_\delta^{nm0} +
%      \frac{2\re[\Delta k_\varrho^{nm0}]}{k_\varrho^{nm}} - \epsilon^{nm0}\right)\frac{\omega^{nm0}_0}{2}\right.\\
%     & \left.+\, \ii  \left( \epsilon^{nm0} - 2\frac{\re[\Delta k_\varrho^{nm0}]}{k_\varrho^{nm}} +
%      \Gamma_\text{c}^{nm0}\right) \frac{\omega^{nm0}_0}{2}
%         \right]^{-1}.
\end{align}
Considering
$\omega_0^{nm0}\approx \omega^{nm0}$ and $k_0^{nm0}\approx k^{nm0}$,
and  using  \eqref{Deltak2}, 
we obtain 
\begin{align}
    a_1^{nm0} &= \frac{\ii }{2k^{0ml}H}
    \frac{ \omega^{nm0}}{\omega-\omega^{nm0}
         +   \tfrac{1}{2}\ii\,\omega^{nm0} \left( \Gamma_\delta^{nm0} + 
     \Gamma_\text{c}^{nm0}\right)}.
     \label{aznm0_final}
\end{align}

Now we turn to the case $l\neq 0$.
After expanding the square root of  \eqref{kz3} for a small argument, the axial wavenumber becomes
\begin{subequations}
    \begin{align}
        \label{kznmlfinal}
        k_z^{nml} (\omega) &= k_{z}^{l}  
        + \Delta k_z^{nml}(\omega),
        \\
        \nonumber
        \Delta k_z^{nml}(\omega) &= \frac{k_0^{nml}\,(\omega-\omega^{nml})}{c_0 k_{z}^l}
        -\frac{\left(k_0^{nml}\right)^2}{2k_{z}^l}\left(\Gamma_\delta^{nml}-\ii \Gamma_\text{c}^{nml}\right)
        -
        \frac{k_{\varrho}^{nm} }{k_{z}^l} \Delta k_\varrho^{nml}
        \\
        &\phantom{=} - \delta_{n,0}\delta_{m,1}
        \frac{\left(\Delta k_\varrho^{01l}\right)^2}{2k_{z}^l}.
        \label{Deltakz}
    \end{align}
\end{subequations}
To derive the $a_1^{nml}$ coefficient, we replace  \eqref{k0nml} and \eqref{kznmlfinal} into~\eqref{a1General},
and expand the result around $\Delta k_z^{nml}H=0$ to the linear approximation,
\begin{equation}
    \label{aznml}
        a_1^{nml} =   \frac{\ii\left(k_0^{nml}\right)^2}{
        k_0^{nml}k_{z}^{l} \Delta k_z^{nml} H + (1+\ii)\,\epsilon^{nml}\, (k_{\varrho}^{nm})^2
        }.
\end{equation}
Substituting   \eqref{Deltakz} into  \eqref{aznml}, we arrive at
\begin{align}
\nonumber
&a_1^{nml} = \\
&\frac{\tfrac{\ii c_0}{H}}{\omega-\omega^{nml} - \frac{\omega^{nml}}{2}
\left[ 
\Gamma_\delta + \frac{2 k_\varrho^{nm} \Delta k_\varrho^{nml} }{ (k_0^{nml})^2 } +
    \frac{\delta_{n,0}\delta_{m,0}\left(\Delta k_\varrho^{00l}\right)^2}{(k_0^{nml})^2}
    -
    \frac{2\epsilon^{nml} (k_\varrho^{nm})^2}{(k_0^{nml})^3H}
    -\ii \left[ \Gamma_\text{c}^{nml} 
    + 
    \frac{2\epsilon^{nml} (k_\varrho^{nm})^2}{(k_0^{nml})^3H}
    \right]
\right] }.
     \label{aznmlfull}
\end{align}
At the resonance ($\omega=\omega^{nml}$), the real part of the denominator of $a_1^{nml}$ is zero.
Thus, we obtain the relation
\begin{equation}
     \Gamma_\delta + \re\!\left[\frac{2 k_\varrho^{nm} \Delta k_\varrho^{nml} }{ (k_0^{nml})^2 } +
    \frac{\delta_{n,0}\delta_{m,0}\left(\Delta k_\varrho^{00l}\right)^2}{(k_0^{nml})^2}\right]
    -
    \frac{2\epsilon^{nml} (k_\varrho^{nm})^2}{(k_0^{nml})^3H}
        = 0.
        \label{confGamma_nml}
\end{equation}
Revisiting  \eqref{Deltakrnml}, we note that $\re[\Delta k_\varrho^{nml}] =
\im[\Delta k_\varrho^{nml}]$
and
$\re[(\Delta k_\varrho^{00l})^2] =
\im[(\Delta k_\varrho^{00l})^2]$.
Therefore, we can express \eqref{aznmlfull} using \eqref{confGamma_nml} as
% \begin{align}
% \nonumber
% a_z^{nml} &= \frac{\rho_0 c_0^2 \omega^{nml}_0 d_0 }{H} \\
% \nonumber
% &\phantom{=}\times
%         \left\{
%         (\omega-\omega^{nml})
%         -\frac{\omega^{nml}_0}{2 }\Gamma_\delta^{nml}
%         +
%         \frac{\epsilon^{nml}\,c_0^2 (q_\varrho^{nm})^2 }{\omega^{nml}_0}
%         -
%         \frac{c_0^2 q_\varrho^{nm} }{\omega^{nml}_0}
%         \re\!\left[\Delta k_\varrho^{nml}\right]
%         \right.
%         \\\nonumber
%         &\phantom{=} -  \delta_{n,0}\delta_{m,1}
%         \frac{c_0^2}{2 \omega^{01l}_0}
%         \re\!\left[\left(\Delta k_\varrho^{01l}\right)^2\right]
%         + \ii
%         \left[
%         \epsilon^{nml}
%         \frac{c_0^2 (q_\varrho^{nm})^2}{\omega^{nml}_\text{N}} 
%         -
%         \frac{c_0^2 q_\varrho^{nm} }{\omega^{nml}_0}
%         \re\!\left[\Delta k_\varrho^{nml}\right]
%         \right. \\
%         &\phantom{=}
%         \left.\left.
%         -  \delta_{n,0}\delta_{m,1}
%         \frac{c_0^2}{2 \omega^{01l}_0}
%         \re\!\left[\left(\Delta k_\varrho^{01l}\right)^2\right]
%         +
%         \frac{\omega^{nml}_0}{2 }\Gamma_\text{c}^{nml}
%         \right]
%         \right\}^{-1}
%         , \quad l>0.
%         \label{aznmlfull2}
% \end{align}
\begin{equation}
a_1^{nml} = \frac{\ii }{k^{nml}H}
\frac{\omega^{nml}}{\omega-\omega^{nml}
         +   \tfrac{1}{2}\ii\,\omega^{nml} \left( \Gamma_\delta^{nml} + 
     \Gamma_\text{c}^{nml}\right)},
     \quad l\neq0.
     \label{aznml_final}
\end{equation}

We proceed to determine the boundary-layer damping factor $\Gamma_\delta^{nml}$ at the resonance.
By combining  \eqref{Deltak2} and \eqref{confGamma_nml}, we obtain
\begin{equation}
    \Gamma_\delta^{nml} =
     \frac{1}{\left(k_0^{nml}\right)^2}\left(
      \frac{A^l\epsilon^{nml} \left(k_{\varrho}^{nm}\right)^2}{k_0^{nml}H} 
     -
     \re\!\left[
     2 k_{\varrho}^{nm}
     \Delta k_\varrho^{nml}
     + \delta_{n,0}\delta_{m,0}
    \left(\Delta k_\varrho^{00l}\right)^2
     \right]
     \right),
     \label{Gammadnml}
\end{equation}
where $A^0=1$ and $A^l=2$ for $l\neq 0$.
Inserting  \eqref{Deltakrnml} 
into this equation results
\begin{equation}
    \Gamma_\delta^{nml} =
     \frac{\epsilon^{nml}}{\left(k_0^{nml}\right)^2}
     \left\{
     \left[\frac{A^l}{k_0^{nml} H}  + 
    \frac{1}{k_0^{nml} R}
    \frac{(k_{z}^l R)^2 + n^2 }{(k_{\varrho}^{nm} R)^2-n^2} 
     \right] \left(k_{\varrho}^{nm}\right)^2
     + \frac{ \delta_{n,0} \delta_{m,0}}{ k_0^{nml} R}
  \left(k_{z}^{l}\right)^2
     \right\}.
     \label{Gammanml}
\end{equation}

\vspace{.5cm}

\noindent
\textbf{Acknowledgments}
\vspace{.1cm}

G. T. Silva would like to thank 
the Brazilian National Council for Scientific and Technological Development--CNPq (Grants number 173511/2023-6, 302485/2023-6, 444099/2023-9).
%the Coordination for the Improvement of Higher Education Personnel--CAPES (Program CAPES/COFECUB, Grant number 88881.711904/2022-01). 

\vspace{.5cm}

\noindent
\textbf{Declaration of interests}
\vspace{.1cm}

The author reports no conflict of interest.

\bibliographystyle{jfm}
\bibliography{refs.bib}

%\bibliography{jfm-instructions}

\end{document}